%% file: main.tex
\documentclass[10pt,twocolumn,letterpaper]{article}

\PassOptionsToPackage{numbers}{natbib}

\usepackage{cvpr}              
\makeatletter
\robustify\@latex@warning@no@line
\makeatother
\usepackage{authblk}
\renewcommand{\Affilfont}{\normalsize}

\makeatletter
\renewcommand\AB@affilsepx{, \protect\Affilfont}
\makeatother
\usepackage[colorlinks=true, urlcolor=blue, linkcolor=red]{hyperref}

\usepackage{graphicx}
\usepackage{amsmath}
\usepackage{amssymb}
\usepackage{booktabs}

\usepackage[utf8]{inputenc} 
\usepackage[T1]{fontenc}    
\usepackage{url}            
\usepackage{amsfonts}       
\usepackage{nicefrac}       
\usepackage{microtype}      
\usepackage{xcolor}         
\usepackage{natbib}         
\usepackage{mathtools}      %
\usepackage{comment}
\usepackage{subcaption}
\usepackage{graphbox}
\usepackage{breqn}
\usepackage[capitalize]{cleveref}
\usepackage{enumitem}
\usepackage{capt-of}

\newcommand{\todo}[1]{}
\renewcommand{\todo}[1]{{\color{red} TODO: {#1}}}

\title{\vspace{-1.9cm}\textsc{AMMeBa}: A Large-Scale Survey and Dataset of Media-Based Misinformation \emph{In-The-Wild}}

\author[1]{Nicholas Dufour \thanks{}}
\author[1]{Arkanath Pathak}
\author[1]{Pouya Samangouei}
\author[1]{Nikki Hariri}
\author[2]{Shashi Deshetti}
\author[3]{Andrew Dudfield}
\author[4]{Christopher Guess}
\author[5]{Pablo Hernández Escayola}
\author[1]{Bobby Tran}
\author[1]{Mevan Babakar}
\author[1]{Christoph Bregler}
\affil[1]{Google}
\affil[2]{Factly Media \& Research}
\affil[3]{Full Fact}
\affil[4]{Duke University Reporters’ Lab}
\affil[5]{Maldita.es}

\date{March 2024}

\begin{document}
\pagenumbering{arabic}

\twocolumn[{%
\renewcommand\twocolumn[1][]{#1}%
\maketitle
\begin{center}
    \centering
    \vspace{-1.0cm}
    \includegraphics[width=1.0\linewidth]{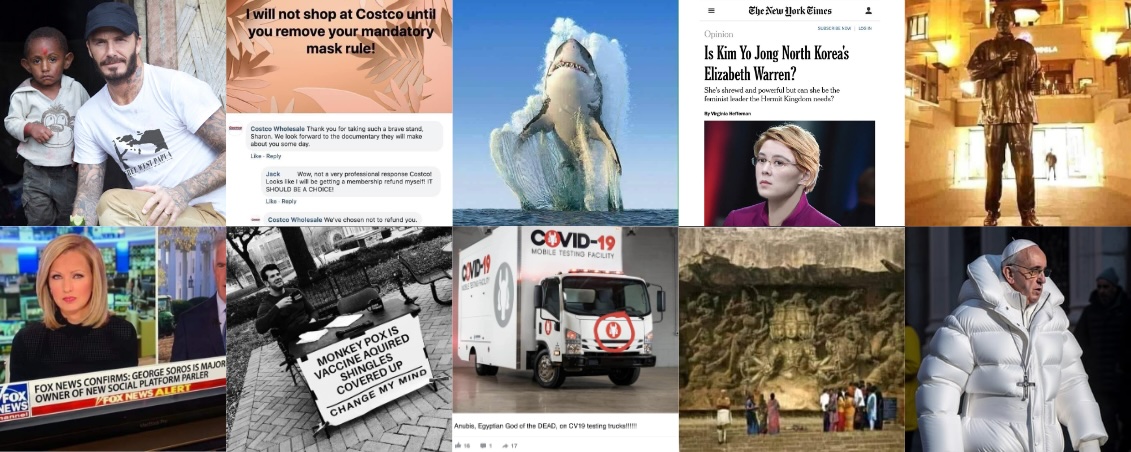}
    \captionof{figure}{\textbf{Examples of media occurring alongside fact-checked misinformation claims}. In this report, we introduce a typology to capture the enormous variation in media-based (particularly image-based) misinformation seen in-the-wild and categorize a very large sample of misinformation claims with it.}
    \label{fig:galileo_examples}
\end{center}%
}]

{
  \renewcommand{\thefootnote}%
    {\fnsymbol{footnote}}
  \footnotetext[1]{Corresponding author: {\tt\small ndufour@google.org}}
}

\begin{abstract}
   The prevalence and harms of online misinformation is a perennial concern for internet platforms, institutions and society at large. Over time, information shared online has become more media-heavy and misinformation has readily adapted to these new modalities. The rise of generative AI-based tools, which provide widely-accessible methods for synthesizing realistic audio, images, video and human-like text, have amplified these concerns. Despite intense public interest and significant press coverage, quantitative information on the prevalence and modality of media-based misinformation remains scarce. Here, we present the results of a two-year study using human raters to annotate online media-based misinformation, mostly focusing on images, based on claims assessed in a large sample of publicly-accessible fact checks with the ClaimReview markup. We present an image typology, designed to capture aspects of the image and manipulation relevant to the image's role in the misinformation claim. We visualize the distribution of these types over time. We show the rise of generative AI-based content in misinformation claims, and that its commonality is a relatively recent phenomenon, occurring significantly after heavy press coverage. We also show ``simple'' methods dominated historically, particularly context manipulations, and continued to hold a majority as of the end of data collection in November 2023. The dataset, \textbf{A}nnotated \textbf{M}isinformation, \textbf{Me}dia-\textbf{Ba}sed (\textsc{AMMeBa}), is publicly-available, and we hope that these data will serve as both a means of evaluating mitigation methods in a realistic setting and as a first-of-its-kind census of the types and modalities of online misinformation. 
\end{abstract}

\section{Summary of Findings}\label{sec:summary}
\input{sections/0_summary}

\section{Introduction}\label{sec:introduction}
\input{sections/1_introduction}

\section{Related Work}\label{sec:related_work}
\input{sections/2_related_work}

\section{Source Data}\label{sec:source_data}
\input{sections/3_source_data}

\section{Media Typology}\label{sec:media_typology}
\input{sections/4_media_typology}

\section{Data Annotation}\label{sec:data_annotation}
\input{sections/5_data_annotation}

\section{Results}\label{sec:results}
\input{sections/6_results}

\section{Conclusion}\label{sec:conclusion}
\input{sections/7_conclusion}

\section{Limitations \& Future Research}
\label{sec:limitations_further}
\input{sections/8_limitations_future}

\begin{footnotesize}
\section{Acknowledgments}

The authors wish to extend a special thanks to the numerous fact checkers whose work is a crucial vanguard against misinformation and forms the basis for \textsc{AMMeBa}. A number of collaborators within Google also helped make this possible: Jason Brown, Alessandro Bucelli, Scott Frohman, Reza Ghanadan, Sudhindra Kopalle, Thomas Leung, Alexios Mantzarlis, Andrew Moore, Tony Neves, Christopher Savçak, and Avneesh Sud. We also thank our external colleagues, especially Annalisa Verdoliva at the University Federico II of Naples as well as the Duke University Reporters' Lab. The authors also wish to thank the team of annotators involved in preparing this dataset. This research was, in part, funded by the U.S. Government. The views and conclusions contained in this document are those of the authors and should not be interpreted as representing the official policies, either expressed or implied, of the U.S. Government.

\end{footnotesize}

\bibliographystyle{plain}
\bibliography{references}

\appendix
\input{sections/appendix}

\end{document}

%% file: sections/0_summary.tex
\begin{itemize}
    \item Misinformation claims from a total of 135,838 fact checks were analyzed dating back to 1995, although the bulk were created after the introduction of ClaimReview in 2016.
    \item A large majority of these claims (most recently, about 80\%) involve media.
    \item Images were historically the dominant modality associated with misinformation claims; however, videos became more common starting in 2022 and now participate in more than 60\% of fact-checked claims that include media.
    \item Despite widespread concern since the late 2010s, AI-generated content was rare until Spring of 2023, when their presence in fact check misinformation claims dramatically increased.
    \item Image manipulations in the annotated claims were generally simple and do not require technological sophistication to produce; the most common type are context manipulations, which use (frequently unmodified) images alongside a false claim about what they depict.
    \item Text is very common in the images (often articulating the misinformation claim itself), occurring over or alongside the visual content of the image.
\end{itemize}

%% file: sections/1_introduction.tex
\begin{figure}
	\centering
    \includegraphics[width=1\linewidth]{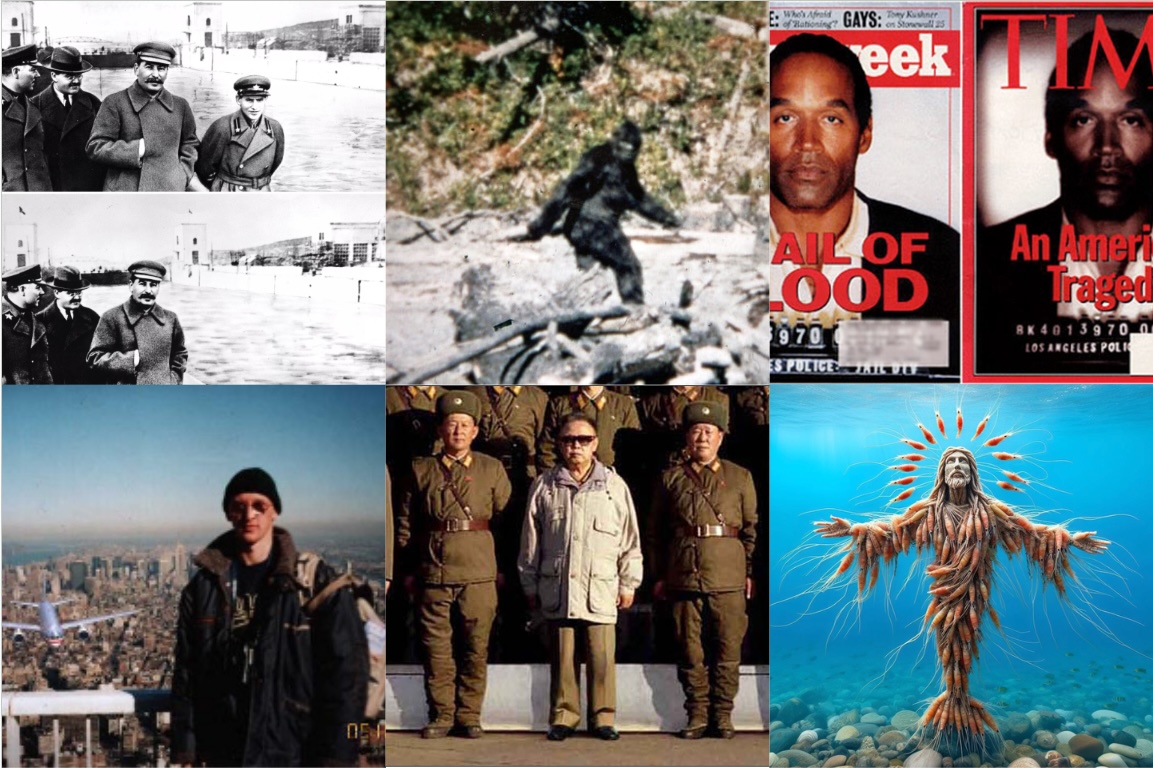}
    \caption{\textbf{Media manipulations have a long history}. \underline{Top Left}: A comparison of an image of Joseph Stalin, originally taken in 1937, where an associate, Nikolai Yezhov, is present along with a later version where he has been manually removed from the official image with airbrushing, following his fall from favor. \underline{Top Middle} A still from the 1967 ``Patterson-Gimlin film,'' alleged to show the cryptid Bigfoot, taken along the Klamath River in California. While it's authenticity is disputed to this day, experts regard the footage as a hoax. \underline{Top Right} A 1994 Time Magazine cover featuring a mugshot of O.J. Simpson that had been artificially darkened to appear more sinister, compare the same photograph on the cover of Newsweek. \underline{Bottom Left} The ``Tourist Guy'' image, originally taken in 1997. The creator of the image confessed to manipulating the image to add the airplane ``as a joke.'' \underline{Bottom Middle} A 2008 image of Kim Jong Il that a BBC analysis concluded was fake. \underline{Bottom Right} A 2024 AI-generated image, dubbed ``Shrimp Jesus,'' that went viral on social media.} 
    \label{fig:historical_fakes}
\end{figure}

With examples dating back to antiquity, mis- and disinformation\footnote{Researchers regard misinformation and disinformation either as distinct categories or consider disinformation to be a type of misinformation, distinguished by it's intent to misinform, whereas misinformation may be inadvertent. We do not collect information to distinguish between these categories, and thus we will refer to these types collectively as \emph{misinformation} for the remainder of this report.} are not new phenomena \cite{posetti2018short} (\cref{fig:historical_fakes}). Concerns about their prevalence and impact, however, have grown more acute in the social media era, where one-to-many distribution channels with global audiences have grown vastly more accessible and popular. This concern is driven by a number of factors both empirical, with studies showing heavy engagement with misinformation on social media platforms \cite{allcott_et_al_2019}, and perceptual, as numerous polls indicate misinformation is a concern among an ``overwhelming'' majority of the public \cite{unesco_ipsos_survey_2023,kff_survey_2023,ap_norc_survey_2023}.

The explosion of generative AI-based methods has inflamed these concerns, as they can synthesize highly realistic audio and visual content as well as natural, fluent text at a scale previously impossible without an enormous amount of manual labor. Such methods are improving rapidly in both the quality of the synthesized content and their adherence to user intents. Further, while formerly the province of AI researchers, generative AI tools are now freely available to those with internet access or the compute resources to run them locally. Their outputs, which are collectively termed ``Deepfakes''\footnote{The term ``Deepfake'' was originally coined in late 2017 on Reddit (https://www.reddit.com/) in a since-deleted post and referred specifically to neural-network media face swaps.} when not text-based, are seen as a means to create and amplify misinformation and potentially produce a number of harms \cite{chesney2019deep}.

Interest in online misinformation has historically focused primarily on text-based claims \cite{wardle2017information}. The presence and character of media that co-occurs with that text has received comparatively less attention. This is somewhat surprising; media use online is widespread: platforms like Youtube\footnote{https://www.youtube.com/} and TikTok\footnote{https://www.tiktok.com/en/} each boast more than a billion users, and are exclusively video-based. Nearly 3.5 billion images were shared daily across social media platforms as of 2016 \cite{meeker2016internet}, a number that is surely higher today. 

The presence of media affects engagement, increases it in at least the case of images \cite{li2020picture}, and images bearing or associated with misinformation have higher engagement than other images \cite{wang2021understanding}. Images shared alongside claims are perceived as relevant and can alter perceptions of that claim's accuracy, the conclusions drawn on the part of the reader,  and even manipulate memories of events in the case of deepfakes \cite{newman2023misinformed}. Images are also regarded as highly persuasive and effective means of messaging \cite{wardle2017information,weikmann2023visual,hameleers2020picture,Birdsell1996}, including when the communication is effectively a lie. A variety of reasons are offered for this, including differences in how visual content is processed relative to text, that visual content might provide an objective ``index'' of reality, and its ``integrative'' role highlighting specific points.

AI-generated images appear to obtain high engagement as well. The Facebook Widely Viewed Content Report for Quarter 3, 2023 \cite{facebook_widely_viewed_content_2023_Q3} includes an AI generated image \cite{diresta2024spammers} on it's list of the top 20 ``Widely viewed posts.'' DiResta \& Goldstein \cite{diresta2024spammers} surveyed more than a hundred Facebook pages that post AI-generated images and find they are highly followed, with user comments indicating that ``many users are unaware of their synthetic origin.'' 

In this study we focus on media-based misinformation, cases where media co-occurs alongside a misinformation claim (or containing the claim itself), and where the claim's effectiveness depends materially on the presence of that media. We hope to provide an accounting of the presence and nature of the media used in such claims, and the manipulations present, that is both large-scale and granular. To this end, we used human raters to annotate claims assessed in fact checks with the ClaimReview\footnote{https://www.claimreviewproject.com/} markup and available on the open web. 

The bulk of the annotations center in particular on misinformation claims that involve images. The characterization, and the associated typology, are oriented around the mechanism by which the images are used to bolster the misinformation claim: the inclusion of text, the type of manipulation, etc. As such, annotations address the types of manipulations performed and how they are realized, rather than the subject matter of the claims themselves, to ensure the information is instrumentally useful to those working in developing tools and mitigation methods that rely on the properties of the media or its context. 

We term the resulting dataset \textsc{AMMeBa}: \textbf{A}nnotated \textbf{M}isinformation, \textbf{Me}dia-\textbf{Ba}sed. \textsc{AMMeBa} annotations are available online through Kaggle \href{https://www.kaggle.com/datasets/googleai/in-the-wild-misinformation-media/}{here}\footnote{https://www.kaggle.com/datasets/googleai/in-the-wild-misinformation-media/}.

%% file: sections/2_related_work.tex
What constitutes misinformation online is very polymorphic, and accordingly, a large variety of surveys have been conducted and datasets collected. Many of these datasets are aimed at researchers who require large-scale data to create and validate computational methods of misinformation mitigation. Many of these are ``boutique'' datasets (\emph{e.g.,} face manipulations in images and video like \textsc{FaceForensics} \cite{Rossler_2019_ICCV}) that focus on a particular topic or domain, although general misinformation datasets do exist. Such datasets vary by modality, whether or not they are derived from in-the-wild data or were constructed explicitly for the purpose of the dataset, and whether they were labeled programmatically or manually, among any number of other features. 

Several datasets focus on fact-checking the claims themselves, rendered in text. The \textsc{Liar} dataset \cite{Wang17j} applies human labeling to short segments from politifact.com, which themselves correspond to claims made in-the-wild, many of which false. A much larger dataset, FEVER \cite{thorne}, consists similarly of claims but is ``synthetic,'' claims were constructed from information available in Wikipedia articles.

The rise of generative AI methods has motivated a number of image-based datasets, which consist of AI-generated images, sometimes with the generator prompt used to obtain the image. Such datasets very frequently not only consist of synthetic data but are also produced in a synthetic way, by direct sampling large numbers of images, for instance \textsc{GenImage} \cite{zhu2023genimage}. Others, however, do capture in-the-wild distributions. Two such datasets collect images and prompts from channels on the chat app Discord\footnote{https://discord.com/} that are specific to popular image generators Midjourney\footnote{https://www.midjourney.com/home} \cite{midjourney_kaggle} and Stable Diffusion\footnote{https://stability.ai/} (\textsc{DiffusionDB}) \cite{wang2023diffusiondb}. While sampled from in-the-wild sources, both restrict their attention to a single synthesizer. TWIGMA \cite{chen2023twigma}, whose samples are inferred to be synthetic based on metadata, uses Twitter\footnote{https://twitter.com/} (now ``X'') as a source and thus aims to be more varied and comprehensive and should emulate the distribution of generators and synthesis methods used online. 

\textsc{PS-Battles} \cite{ps_battles} does not single out a specific manipulation type like generative AI, and obtains samples from a popular sub-reddit\footnote{https://www.reddit.com/r/photoshopbattles/} whose content consists of images manipulated to humorous effect by users in a friendly ``battle.''. However, none of these are sampling specifically from media associated with misinformation or false claims, and as such may not replicate the distribution of image content seen in those settings. Further, extant image-based datasets are generally not useful for understanding prevalence and trends as they don't intentionally capture longitudinal data. Reis \emph{et al.} \cite{reis2020dataset} restricts their attention to images and, like this report, use fact checks as a starting point. They identify when images that are associated with fact-checked claims are shared on popular messaging app WhatsApp in public groups during the leadup to the 2018 Brazilian and 2019 Indian elections. Reis \emph{et al.}'s dataset is more likely than \textsc{PS-Battles} to be representative of misinformation-associated images, but does not provision annotations for the type of manipulation involved.

Misinformation often depends on the interaction between text and media. COSMOS \cite{cosmos} provides image and text pairs and synthetically create ``context manipulations'' (\ref{sec:typology_context_manip}). McCrae \emph{et al.} \cite{mccrae2022multi} construct a similar dataset but using videos instead of images, collected from Facebook\footnote{https://www.facebook.com/}. McCrae \emph{et al.} use real (video, caption) pairs as authentic examples, but like COSMOS construct synthetic manipulations by permuting those pairs. These satisfy the technical definitions of the context manipulations (which are often deployed in misinformation, \emph{see} \cref{sec:context_manip}) they seek to model but they lack the intent and bespoke curation seen in authentic samples.

Datasets exist that attempt to capture a wider scope, with no restriction to a specific modality, manipulation type, or synthesis method. \textsc{r/Fakeddit} \cite{nakamura2020rfakeddit} uses posts from a variety of sub-reddits, entailing all modalities that can be included in Reddit posts. However, \textsc{r/Fakeddit} labels are relatively low granularity, and are inferred based on the source sub-reddit. \textsc{MuMiN} \cite{nielsen2022mumin} is even broader; like this study, \textsc{MuMiN} starts from a large number of fact checks, extracting claims and then finding associated content. However, little additional structured information is provided beyond the simple fact that the content and claim are associated. 

While datasets provide subject matter exemplars (either empirical and in-the-wild or synthetic), \textbf{surveys} make structured claims about prevalence, subject, modality or type of online misinformation. As this report is a blend of the two, we also briefly review preceding work done in this area. The enormity of misinformation as a subject means such surveys are often specific, constrained by platform, time or subject matter. There are a number of studies that examine misinformation in a topic-specific way (\emph{e.g.}, COVID-19, 5g cell towers, Q-Anon). Brennan \emph{et. al} \cite{brennen2020types} is one such example; focusing on COVID-19 misinformation, they sample misinformation from fact checks and provide granular annotations. Like this study, they find a significant contingent of media-based misinformation is context-based manipulations, with content manipulations making up a significantly smaller proportion. 

An alternative, less content-specific approach is to restrict analyses to a specific platform. Grinberg \emph{et al.} \cite{grinberg2019fake} study the spread of misinformation on Twitter leading up to the 2016 United States presidential election, focusing primarily on the frequency of exposure to fake news by Twitter users, particularly as a function of political affinity, demographics, and consumption of political content overall. Addressing visual misinformation specifically, Yang \emph{et al.} \cite{yang2023visual} manually annotated a random sample of political images posted to Facebook, and find a large contingent contain misinformation elements. Yang \emph{et al.} additionally report ``four main types'' of misleading image posts (doctored images, memes with misleading text, images associated with misleading posts, and text-only images). These categories map fairly neatly onto those outlined in this study. However, they do not provide further details beyond this enumeration.

%% file: sections/3_source_data.tex
Misinformation claims were sampled using publicly-available fact checks with the ClaimReview markup, which is used by 60\% of fact checkers surveyed by IFCN in 2023 \cite{StateoftheFactCheckersReport2023}. In total, 135,862 English-language fact checks were annotated at some level. The limitations on this study imposed by the restriction to English-language fact checks, and the use of fact checks in general as a source of in-the-wild misinformation claims is discussed in \cref{sec:lims_fut_source_data,sec:lims_fut_language}. 

\begin{table}[h]
\centering
\begin{tabular}{|l|l|}
\hline
\multicolumn{2}{|c|}{\textbf{Fact check counts}} \\
\hline
\textbf{Webpage} & \textbf{Fact Check Count}\\\hline
snopes.com & 15,751 \\\hline
politifact.com & 14,241 \\\hline
checkyourfact.com & 9,652 \\\hline
factcheck.afp.com & 7,623 \\\hline
leadstories.com & 7,623 \\\hline
fullfact.org & 5,441 \\\hline
boomlive.in & 5,310 \\\hline
newsmobile.in & 4,776 \\\hline
factly.in & 4,535 \\\hline
misbar.com & 4,135 \\\hline
\end{tabular}
\caption{\textbf{Fact check counts by website.} Fact check websites, ranked by the number of annotated fact checks in this study. Fact checks from a total of 166 unique fact checker webpages were subject to annotation.}
\label{table:fact_check_by_domain}
\end{table}

\begin{figure}[h]
    \centering
    \includegraphics[width=0.9\linewidth]{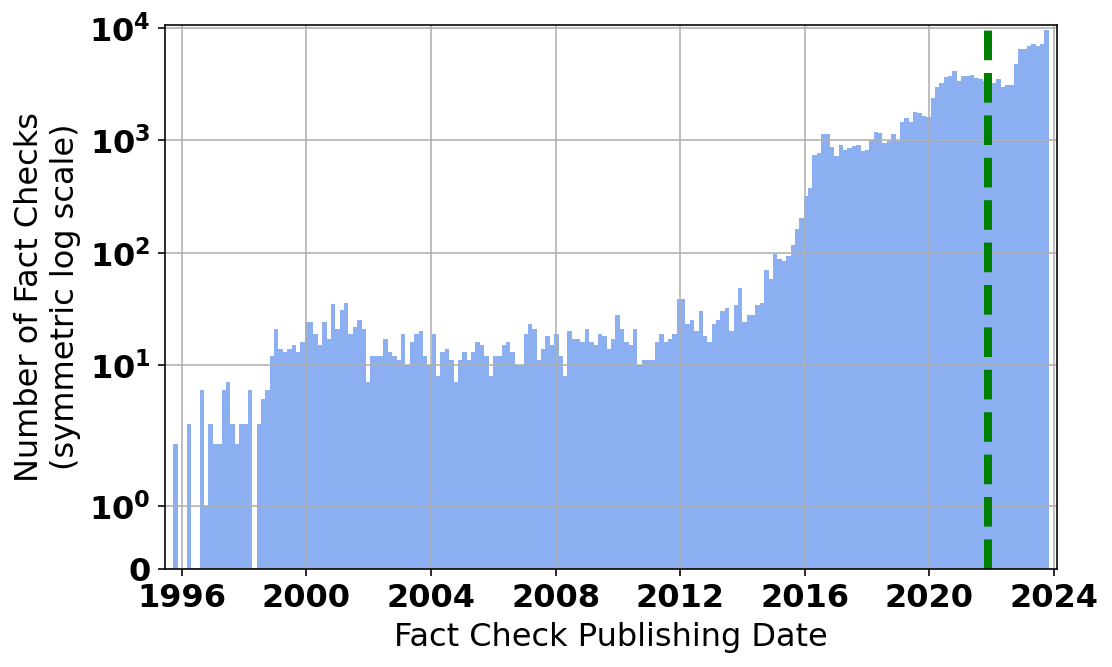}
	\caption{\textbf{Count of fact checks with annotations by publication date}. Dates are approximate, and rely on the date reported by the fact checker or the date where the fact check was first encountered, preferring the self-reported dates where available. The dashed green line indicates the start of data collection, November 11, 2021.}
\label{fig:fact_check_pubdate}
\end{figure}

Fact check publication dates range from 1996 to the end of data annotation on November 18, 2023. A histogram of counts arranged by date is shown in \cref{fig:fact_check_pubdate}. The majority of fact checks annotated are recent, with far fewer prior to 2016, which accords with the schema's introduction in March of 2016 (ClaimReviews for older fact checks were added \emph{post hoc}).

In total, we sampled fact checks from 166 domains from fact checking organizations worldwide. The top 10 domains with the largest numbers of fact checks present in the annotation dataset is shown in the \cref{table:fact_check_by_domain} along with their counts. 

During annotation, 633 fact checks were found to be removed or otherwise no longer accessible. These fact checks were not subject to further annotation, although any prior annotations were retained.

%% file: sections/4_media_typology.tex
Media were classified by raters (\cref{sec:dataset_raters}) according to one of several typologies, which reflect aspects of the media related to forensics and provenance, rather than subject matter. The objective of this approach is to capture attributes relevant to understanding and identifying trends in how online media-based misinformation is created rather than trends in misinformation narratives. We hope this will help those working to combat misinformation prioritize research directions and better understand the broad landscape as it relates to the development of novel methods. Below, we outline these typologies and provide examples. Further details, particularly the reasoning and motivation for their structure, are available in the Appendix (\ref{sec:appx_media_typology}). 

\subsection{Top Level Types} \label{sec:typology_top_level_annotations}

Fact checks were treated as corresponding to individual misinformation claims. Fact checks that violate this were discarded (\ref{sec:appx_inclusion_criteria}). The corresponding claims were classified broadly into \emph{media-based} or \emph{non-media-based}:
\begin{itemize}
    \item \textbf{Media-based claims} were defined based on the ``material relevance'' of associated media to the claim. Media were materially relevant to the claim if their removal would mean that the claim became incoherent, significantly less effective, lost evidence for its veracity, or entirely absent.
    \item \textbf{Non-media-based claims} either did not have any media present, or the media could be removed or ignored without any consequence to the claim. This is true in cases where the media are unrelated, or only serve an aesthetic purpose, etc.
\end{itemize}

This is a partially subjective judgment; consensus was reached after several rounds of trial labeling and review. 

\subsection{Media-Based Misinformation Claims} \label{sec:typology_media_based}

Media-based misinformation claims were further subdivided according to the modality of media deployed: either \textbf{image}, \textbf{video}, or \textbf{audio}\footnote{Claims that depended on a video that had no visual content (\emph{i.e.,} a blank screen) were classified as audio-based.}-based. Claims that relied on more than one media item (possibly of mixed modality) were labeled as \textbf{multiple-media}-based claims. Subsequent categories focus primarily on images (\ref{sec:lims_fut_modality})

\subsection{Image Types} \label{sec:typology_image_types}

Images were classified according to rough properties of their content, particularly when the content type is known to affect the performance of forensic or provenance-recovery tools (\emph{i.e.,} reverse image search) and whether or not the images bore text content that can be extracted using automated methods and analyzed separately. More details on the considerations that led to the image typology are available in the appendix (\ref{sec:appx_typology_image_types}). Image types were defined carefully and precisely; raters used these criteria to assign image type labels and did not rely on fact check content for categorization.

\begin{figure}
	\centering
    \includegraphics[width=.23\textwidth]{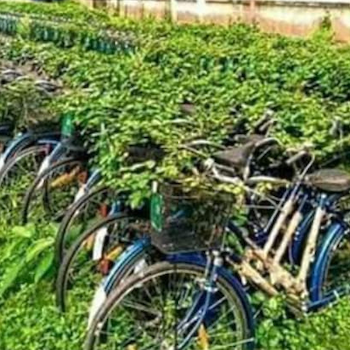}\hfill
    \includegraphics[width=.23\textwidth]{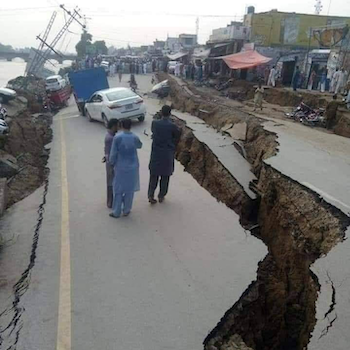}
    \\[\smallskipamount]
    \includegraphics[width=.23\textwidth]{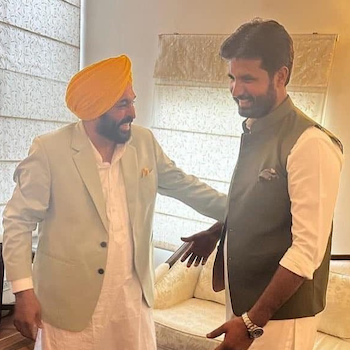}\hfill
    \includegraphics[width=.23\textwidth]{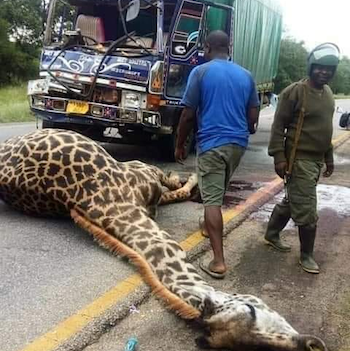}
    \caption{\textbf{Examples of ``basic'' images}. Basic images are photograph-like, although they may be illustrations. They do not contain significant overlaid text, graphical or GUI elements. They are ``coherent,'' depicting a unified scene, and do not comprise multiple sub-images in a collage or mosaic. Note these images have been cropped to square aspect ratio and resized.} 
    \label{fig:basic_image}
\end{figure}

\subsubsection{``Basic'' Images} \label{sec:typology_basic_images}
The first image type category analyzed were termed ``basic images.'' These images appear photographic, although they can include single artworks and synthetic images produced using generative AI or 3D modeling software. They are ``photographic'' in that they represent a coherent scene. They do not include graphical overlays, charts, figures, or images with added digital text. Images containing graphical user interface (GUI) elements indicative of a screenshot are similarly excluded. Finally, images that are digital mosaics or digital collages of multiple images are excluded, even if their constituent components are themselves basic images\footnote{This category includes a number of edge cases that are somewhat ill-defined. For example, a digital image of a well known pre-existing artwork using décollage could be considered a basic image. These cases are very rare.}. Basic images \emph{are} permitted to include small overlays or watermarks, but their presence is annotated by the rater. Examples of basic images are presented in \cref{fig:basic_image}.

\begin{figure}
	\centering
    \includegraphics[width=.23\textwidth]{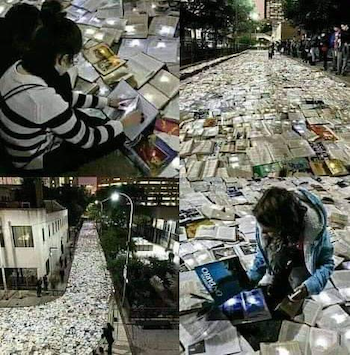}\hfill
    \includegraphics[width=.23\textwidth]{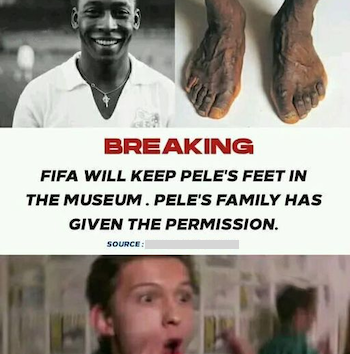}
    \\[\smallskipamount]
    \includegraphics[width=.23\textwidth]{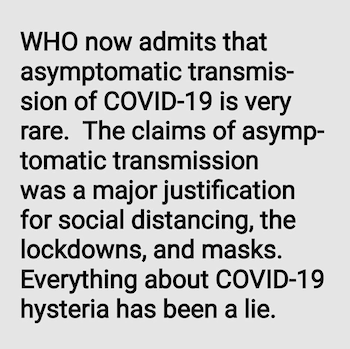}\hfill
    \includegraphics[width=.23\textwidth]{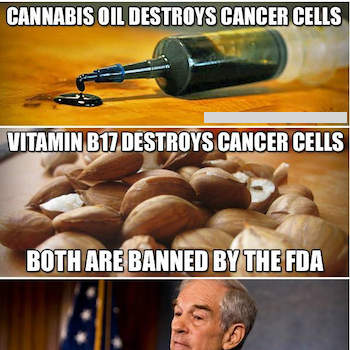}
    \caption{\textbf{Examples of ``complex'' images}. Complex images include all images that are not basic images (\cref{fig:basic_image}, \cref{sec:typology_basic_images}), because they include graphical components, compositions of basic images, or are screenshots. Note these images have been cropped to square aspect ratio and resized.} 
    \label{fig:complex_image}
\end{figure}

\subsubsection{``Complex'' Images} \label{sec:typology_complex_images}
We adopt the term ``complex'' image to identify any image that does not fall into the basic image category, and as such vary considerably as a category. They are principally characterized by the presence of graphical elements (figures, charts, overlay text, symbols, or other elements that are clearly added after-the-fact) or by comprising multiple sub-images. Complex images include screenshots, although these are annotated as a distinct sub-category. Complex images are presented in \cref{fig:complex_image}. Complex images may, and frequently do, include basic images as sub-images. For instance, a montage of basic images is a complex image, as is a screenshot of a social media post that includes a basic image.

\paragraph{Screenshots} \label{sec:typology_screenshots}

Screenshots are a class of complex images that include GUI elements that conclusively show the image was captured by a user taking a screenshot of content on an electronic display. To qualify for this category, the images must contain GUI elements that are unambiguous and recognizable. While a screenshots are a type of complex image, they are broken out here because of their prevalence and the differences in the investigatory approach compared to general complex images. 

Among screenshots, three subtypes are recognized in this study. As screenshots are identified based on the presence of recognizable GUI elements, these subtypes depend on the nature of these GUI elements in addition to descriptions in the body of the fact check. Elements that correspond to social media platforms are classified as \textbf{social media screenshots} while those that have GUI elements from a source other than a social media network are labeled \textbf{non-social media screenshots}. Social media screenshots described in the fact check as fake (\emph{e.g.,} produced by online screenshot generators) are labelled as \textbf{fake social media screenshots}.

\subsubsection{``Analog Gap'' Images} \label{sec:typology_analog_gap}

\begin{figure}
	\centering
    \includegraphics[width=.23\textwidth]{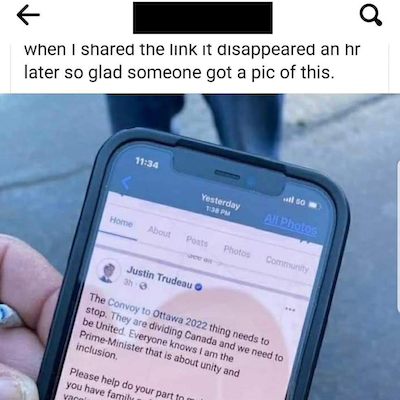}\hfill
    \includegraphics[width=.23\textwidth]{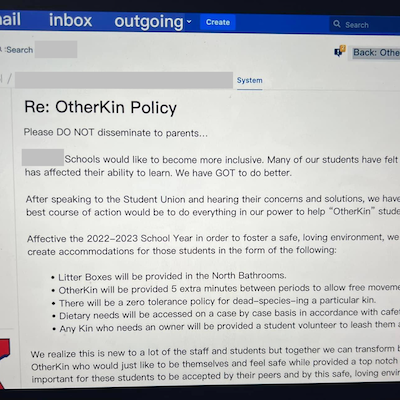}
    \\[\smallskipamount]
    \includegraphics[width=.23\textwidth]{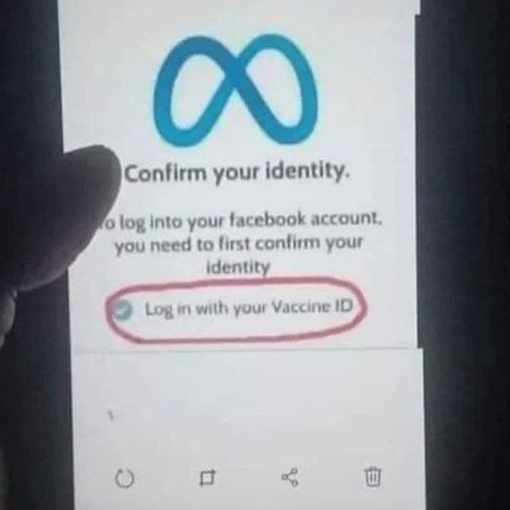}\hfill
    \includegraphics[width=.23\textwidth]{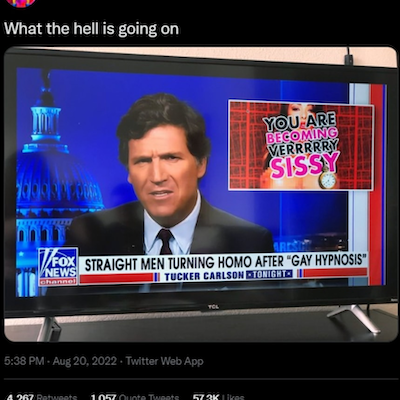}
    \caption{\textbf{Examples of ``Analog Gap'' images}. Analog Gap images occur when misinformation-relevant content is displayed on a screen in the image itself. The image in the top-left is of particular note: this was shared as a screenshot of a Facebook post containing an image of a screen. This is characteristic of misinformation-relevant images online, where successive re-sharing can layer additional elements onto the image, much like sedimentation in the real world. Note these images have been cropped to square aspect ratio and resized. Information identifying the user who posted the content has been removed.} 
    \label{fig:analog_gap_examples}
\end{figure}

We recognize one additional content-dependent image category, ``Analog Gap,'' that is orthogonal to the others. Images are given this label by raters when they (i) feature a screen or a display that (ii) contains content relevant to the misinformation claim. While this category may seem curiously specific, it is recognized separately because media that has been captured in this way is a known weakness of forensic methods (\ref{sec:appx_typology_analog_gap}). Examples of Analog Gap images are displayed in \cref{fig:analog_gap_examples}.

\subsection{Manipulation Types} \label{sec:typology_manip_types}

The \emph{manipulation type} label indicates the specific method used to cause the image to present or support misinformation. We adopt three top-level categories, which are not mutually exclusive:

\begin{itemize}
  \item \textbf{Content manipulations} occur when the pixels of the image itself are altered (or entirely synthesized in the case of generative AI) to change its semantic content in support of the misinformation claim. This category includes manipulation types that generate significant popular attention, like Deepfakes and ``Photoshops.''
  \item \textbf{Context manipulations} occur when false details are provided about the image, \emph{e.g.}, when or where it was taken or what it depicts. Context manipulations often have no content manipulations present: image is presented in an unmodified form alongside the false claim about it. 
  \item \textbf{Text-based images} do not require the visual content of an image to possess any specific qualities. They are misinformation claims, as text, rendered in an image. The image may feature other content, but that content is not important for the claim being made. Images which feature no misinformation-relevant non-textual visual content at all are a common element of this category.
\end{itemize}

\subsubsection{Content Manipulations} \label{sec:typology_content_manip}

\begin{figure}
  \centering
    \includegraphics[width=.23\textwidth]{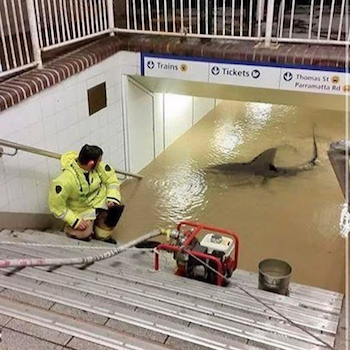}\hfill
    \includegraphics[width=.23\textwidth]{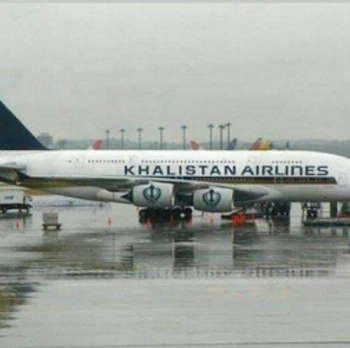}\hfill
    \\[\smallskipamount]
    \includegraphics[width=.23\textwidth]{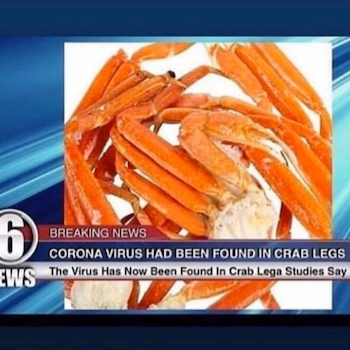}\hfill
    \includegraphics[width=.23\textwidth]{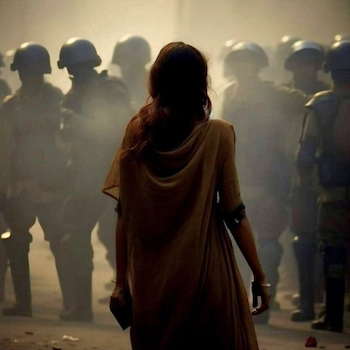}\hfill
    \caption{\textbf{Examples of content-manipulated images}. Content manipulations require that a component of the image itself has been altered in a way that creates or supports a misinformation claim and appears ``native'' to the image. All subcategories are represented here. \underline{Top Left}: A flooded subway, where a shark has been added digitally. This is an example of a general or generic content manipulation, using typical photo manipulation software. \underline{Top Right}: An example of manipulated text, where text that appears to be present in the scene itself (as opposed to overlaid) is added or altered. Text on the airplane originally read ``Singapore Airlines.'' \underline{Bottom Left}: Manipulated news chyron. Unlike manipulated text images, the text does not appear to be present on a real object but added digitally. However, due to its placement on a news chyron, it still appears ``native'' to the image. \underline{Bottom Right}: An example of an image synthesized in its entirety using generative AI. Note these images have been cropped to square aspect ratio and resized.} 
    \label{fig:content_manip_examples}
\end{figure}

Content manipulations are defined here as taking place when (i) the content of the image itself has been altered or synthesized, (ii) it appears endogenous or ``native'' to the image itself (\emph{i.e.}, not clearly overlaid or added after-the-fact\footnote{This is a generous criteria; even very crude attempts to make the manipulations appear present in the original are included here.}) and (iii) the alteration made is relevant to the misinformation claim. Content manipulations have existed since the inception of photography, and are perhaps the canonical form of image-based misinformation in popular consciousness. They are created using a very wide variety of techniques, which are loosely situated onto a spectrum between Deepfakes (\emph{i.e.,} AI-generated) and Cheapfakes (``classical,'' pre-AI methods) \cite{paris2019deepfakes}. 

We asked raters to discriminate between four types of content manipulation. These categories were determined by a combination of forensic relevance, their identifiability given typical fact check content, and rater comprehension. Examples of these manipulations are provided in \cref{fig:content_manip_examples}.

Content manipulation types
\begin{itemize}
  \item \textbf{Manipulated text} occurs when text that appears to be present in the scene itself (\emph{e.g.,} on an object) has been added or altered in a misinformation-relevant way. Because text frequently occurs on saturated, low-information image regions, manipulated text can be more difficult to detect than general content manipulations \cite{chen2008determining}.
  \item \textbf{Manipulated chyrons} include the use of fabricated news chyrons, also called lower-thirds, in images to give the appearance of a screen capture authentic news broadcast to convey misinformation. Unlike manipulated text, chyron text is obviously digital and a viewer recognizes it wasn't present in the underlying image; however, its format as a news broadcast-like overlay allows it to retain authenticity.
  \item \textbf{AI image} are those generated, in whole or in part, by generative AI. Like other content manipulation methods, generative AI encodes forensic signatures that can be detected; however, they are distinct from those in general content manipulations and so are broken out as a distinct category.
  \item \textbf{General} manipulations are content manipulations that don't fit into any of the above categories, such as so-called ``Photoshops.''
\end{itemize}

\subsubsection{Context Manipulations} \label{sec:typology_context_manip}

\begin{figure}
    \centering
    \includegraphics[width=.23\textwidth]{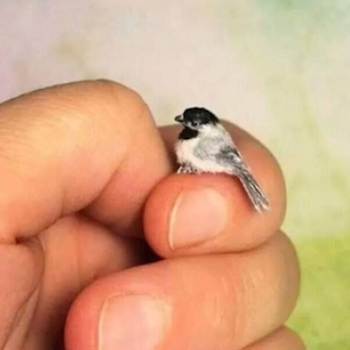}\hfill
    \includegraphics[width=.23\textwidth]{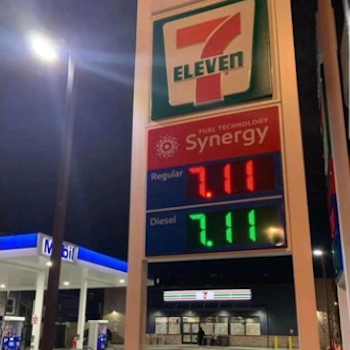}\hfill
    \\[\smallskipamount]
    \includegraphics[width=.23\textwidth]{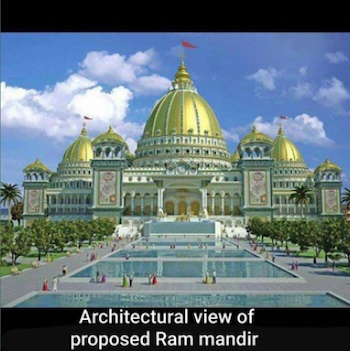}\hfill
    \includegraphics[width=.23\textwidth]{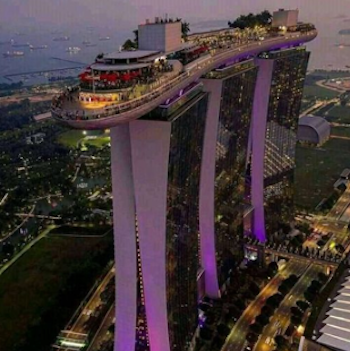}\hfill
    \caption{\textbf{Examples of context-manipulated images}. Context manipulations occur when an image is associated with a claim that provides a false context to the image. \underline{Top Left}: Claimed to be an image of a Zunzuncito ``the smallest bird in the world.'' While the Zunzuncito (\emph{Mellisuga helenae}, a type of tropical hummingbird) is indeed the smallest known species of bird, this is an image of a Chickadee figurine. \underline{Top Right}: Claimed to show accurate gas prices; the display is actually in a testing mode. \underline{Bottom Left}: Claimed to show an ``architectural view'' of the (then in-progress) Ram Mandir; actually shows a visualization of the Temple of Vedic Planetarium in Mayapur, West Bengal. This is an example of an image that provides its own false context, via overlayed text. \underline{Bottom Right}: Claimed to show ``the Tower One Hotel,'' allegedly located in Nakuru, Kenya; actually shows the Marina Bay Sands resort in Singapore. Note these images have been cropped to square aspect ratio and resized.} 
    \label{fig:context_manip_examples}
\end{figure}

Image context manipulations take place when an image is paired with a claim that makes a false assertion about what an image depicts, its origin or its nature. In context manipulations, the image itself is often totally unmodified; it is the accompanying claim that is false. Examples of context manipulations are given in \cref{fig:context_manip_examples} along with the misleading contexts in the caption.

We characterize context manipulations according to the specific aspect of the context that is manipulated. In all of these cases, the manipulation must be relevant to the misinformation rather than incidental. The following categories are not mutually exclusive:
\begin{itemize}
  \item \textbf{Date/Time manipulations}: the claim misleads about the date or time when the image was created. 
  \item \textbf{Location manipulation}: the claim misleads about the location the images was captured or produced.
  \item \textbf{Identity manipulation}: the claim misleads about the identity of an individual or object in the scene depicted.
  \item \textbf{Circumstance manipulation}: the claim mis-characterizes the circumstances of the events in the image, for instance by claiming that an image of a celebration is actually an image of a riot.
  \item \textbf{Atypical manipulation}: claims that omit necessary context or which misrepresent details about the creator of an image (\emph{e.g.,} asserting that fan art is an authentic still from a popular franchise) in a misinformation-dependent way.
\end{itemize}

\paragraph{Self-Contextualizing Images} \label{sec:typology_self_contextualizing_images}

An additional subtype of context manipulations is included in the rater annotations, irrespective of the other context manipulation subtypes and unrelated to the nature of the context manipulation itself. These are ``self-contexualizing images,'' where the false context is provided by text overlaid on the image itself (as opposed to provided by an associated image caption or external statement). These are unique in that they (i) involve a modification to the image, although not one that a viewer would reasonably think was present originally and (ii) do not require that the anyone sharing the content copy the claim themselves in addition to the image: the false claim is embedded in the image content. See \cref{sec:appx_self_contextualizing_image} for further details.

\subsubsection{Text-based images} \label{sec:typology_text_based_image}

Misinformation-associated images do not need to contain photographic or representational content for the association to occur. While not a manipulation in the classical sense, images that make or imply misinformation claims using text present in the image are recognized as a separate manipulation category. Members of this category may have other visual content, but this is unrelated to the misinformation claim itself, which is entirely due to digital text present in the image. This text is clearly superimposed, it is not made to appear as though occurring on an object in the image: they are misinformation claims stated outright.

\subsubsection{Fake Official Documents} \label{sec:typology_false_official_documents}

One additional manipulation category is recognized, the \textbf{``fake'' official document}.

\begin{figure}
    \centering
    \includegraphics[width=.15\textwidth]{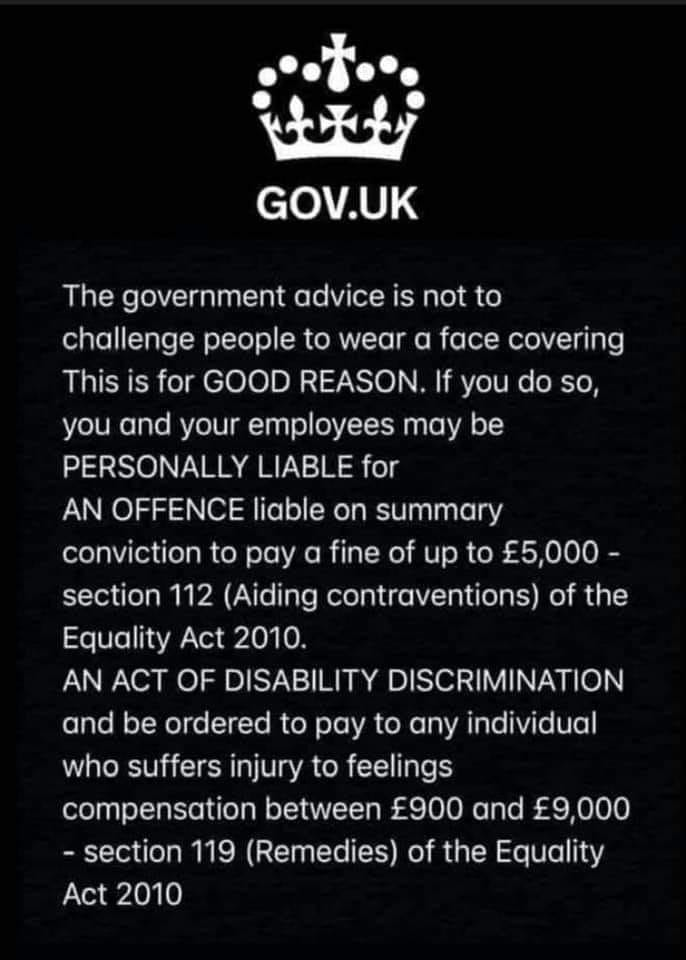}\hfill
    \includegraphics[width=.15\textwidth]{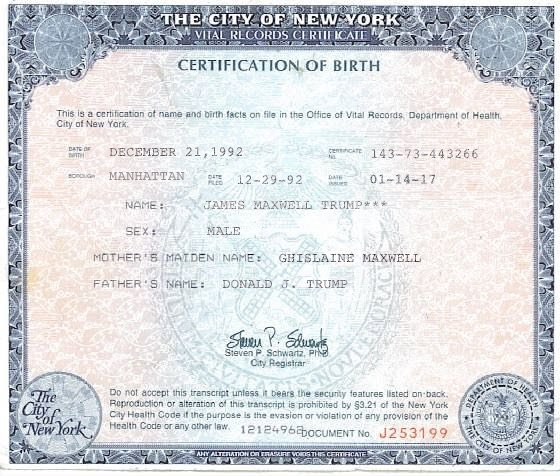}\hfill
    \includegraphics[width=.15\textwidth]{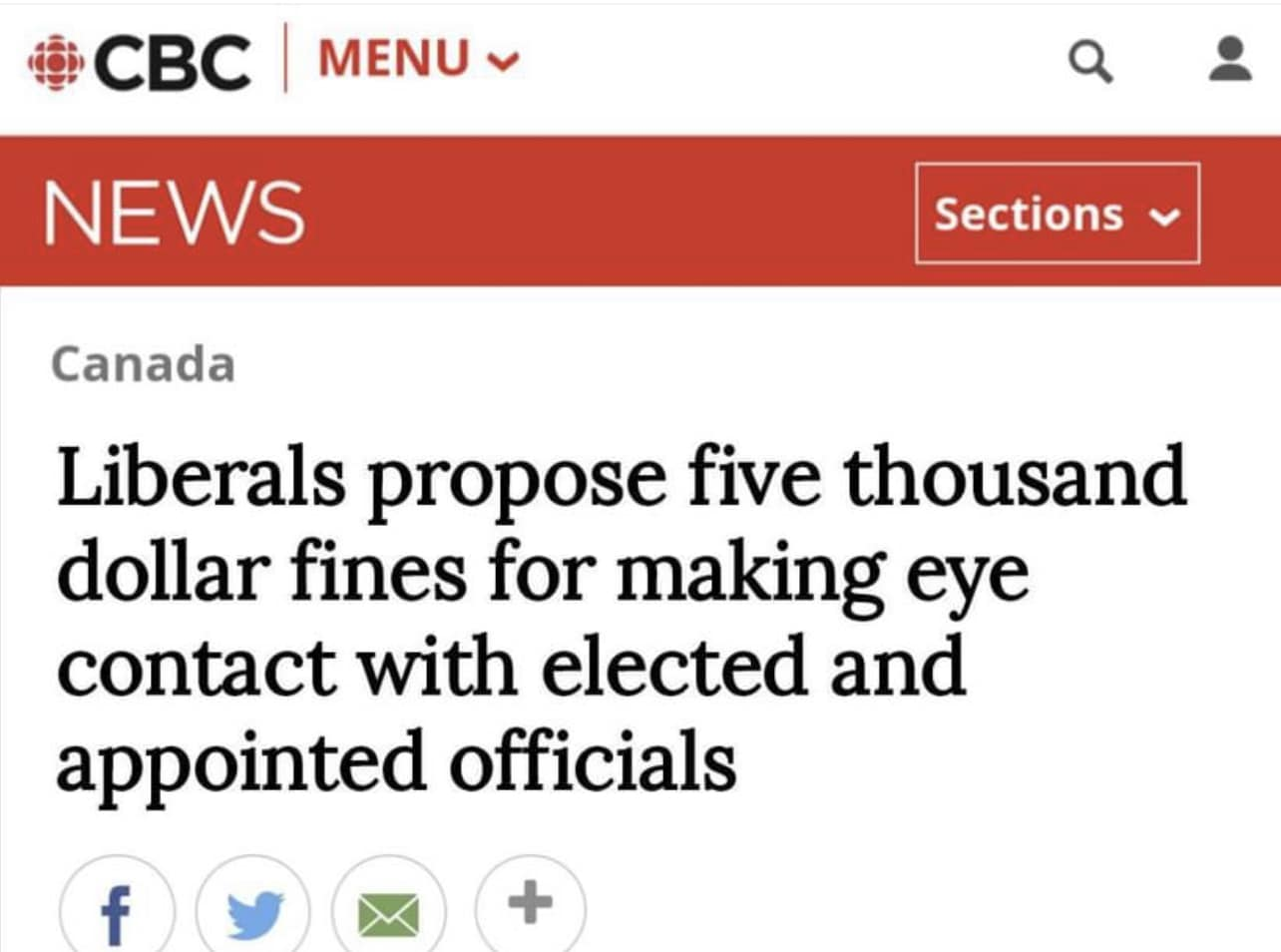}\hfill
    \caption{\textbf{Fake Official Documents}. These images are made to appear as though originating from a trusted, well-known source.} 
    \label{fig:false_official_document_examples}
\end{figure}

This category is characterized by the presence of logos, letterhead, or other attributes that lend the false appearance of an official communication by a government, company, or other well-known organization or official body. They may also appear as articles from reputable news sources, either as physical news clippings or screenshots. Examples of fake official documents are given in \cref{fig:false_official_document_examples}. More detail on this category is available in the appendix (\cref{sec:appx_typology_false_official_documents}).

%% file: sections/5_data_annotation.tex
Data acquisition was a complex endeavor that was subject to extensive refinement during collection.  

\subsection{Task}\label{sec:dataset_task}

\begin{figure}[h]
    \centering
    \includegraphics[width=0.9\linewidth]{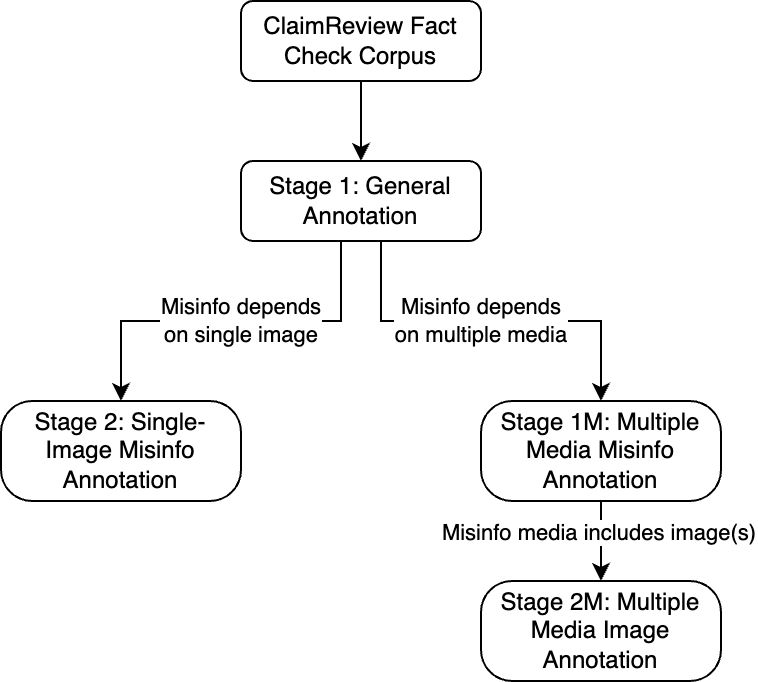}
	\caption{\textbf{Data annotation stage schematic}. A graphical representation of data collection stages. Misinformation claims, as represented by fact checks, were routed through sequential stages based on rater responses in upstream stages.}
\label{fig:task_stages_schematic}
\end{figure}

Annotations were performed via a web interface. The raters were instructed to treat the fact check as a source of truth, and refrain completely from making their own editorial judgements and inferences.

\begin{table*}[t]
\centering
\begin{tabular}{|l|l|l|l|l|}
\hline
& \textbf{Stage 1} & \textbf{Stage 1M} & \textbf{Stage 2} & \textbf{Stage 2M}\\\hline
Collection Begins & Nov 11, 2021 & Jun 16, 2023 & Feb 23, 2023 & Aug 29, 2023 \\\hline
Collection Ends & Oct 31, 2023 & Nov 17, 2023 & Nov 17, 2023 & Nov 18, 2023 \\\hline
Claims Annotated & 135,838 & 5,112 & 32,320 & 14,282 \\\hline
Total Annotations & 249,293 & 37,539 & 83,756 & 27,240 \\\hline
Mean Replication & 1.8 & 7.3 & 2.6 & 1.9 \\\hline
Claims With >1 Annotation & 46.3\% & 98.1\% & 84.9\% & 68.2\% \\\hline
\end{tabular}
\caption{\textbf{Claim and annotation counts by task.} Counts do not include discarded annotations. Stage 2 and Stage 2M are applied to (claim, image) pairs, and as such ``Claims Annotated'' refers to these pairs rather than the number of underlying claims.}
\label{table:task_statistics}
\end{table*}

To reduce cognitive load on the raters and improve throughput, the task was split into four stages, each with different focuses or applied to different subsets, where misinformation claims present in fact checks passed from upstream stages to downstream stages according to upstream annotations. The four stages completed are visualized in \cref{fig:task_stages_schematic}, and statistics about the annotations per stage are presented in \cref{table:task_statistics}. The specific stages are expanded upon in greater detail in \cref{sec:appx_dataset_task}. 

Stages 2 and 2M were focused on fine-grained image annotations. Fact checks were only presented in these stages if the rater in Stage 1 or 1M indicated that (1) an ``original source'' (\emph{i.e.}, the misinformation claim and associated material as it originally existed online and could be encountered by users of the web) was presented by the fact check and could be accessed and (2) that an image was associated with the misinformation claim and available on the original source. Misinformation-associated images that are presented on the fact check page (as opposed to "\emph{in situ}" on the original source) were not considered, as they are often modified by the fact checker to indicate they're fake. Both original web pages and functioning archive versions of those pages (in both cases, they must be linked to by the fact check) were considered original sources.

\subsection{Raters}\label{sec:dataset_raters}

A total of 83 raters participated in the study. Because of the complexity of the task, careful training and retention was prioritized over a large rater pool. Mean rater tenure was 147 days, performing an average of 5,971 annotations (counting annotations that were subsequently dropped or excluded from the analysis due to changes methodology). Two groups of raters were used simultaneously, one located in India and one in the United States. The groups were preferentially allocated fact checks sourced from their respective countries so that annotations would benefit from contextual knowledge; fact checks from neither country were allocated randomly.

%% file: sections/6_results.tex
\subsection{Use of Media in Misinformation Claims}
We hypothesized that media is a pervasive feature of misinformation as it exists in-the-wild, and growing over time as web content continues to become more media-rich.

\begin{figure}[h]
    \centering
    \includegraphics[width=1\linewidth]{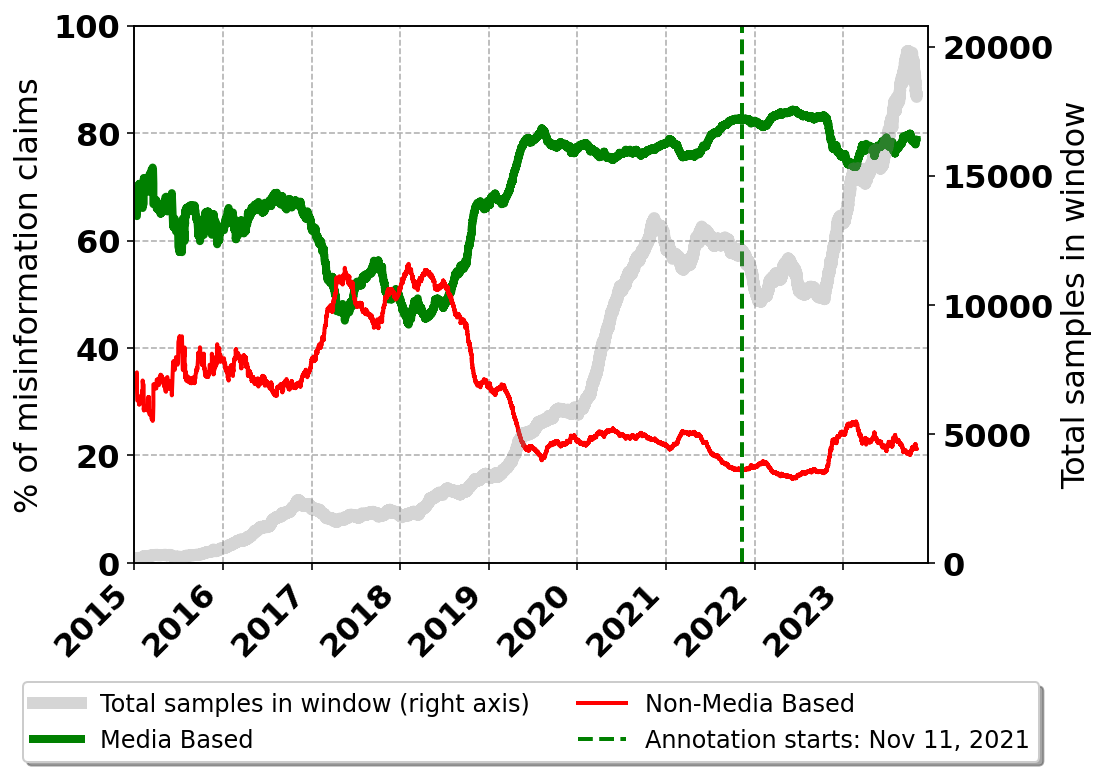}
	\caption{\textbf{Prevalence of media in misinformation claims}. The percentages of media-based and non-media-based misinformation claims may not precisely sum to one due to the presence of ambiguous cases indicated by raters. Percents are calculated according to a 120-day sliding window; the total number of claims present in each window is plotted in gray using the right-hand axis. The dashed green line indicates the start of data collection, November 11, 2021. Dates used for plotting are approximate.}
\label{fig:perc_media_based_claims}
\end{figure}

\cref{fig:perc_media_based_claims} plots the percentage of misinformation claims that rely on media using a 120-day sliding window based on an approximate publication date. While the percentage of media-based misinformation claims among all misinformation claims has been largely stable at least since 2019, such claims represent a large majority of around 80\%, at least among those that are subject to a fact check. The cause of the decline between 2017 and 2019 is unclear; and may be due to changing prioritization in content addressed by fact checkers. The date of the fact check is used to reckon the date of the misinformation claim in this plot, and all subsequent plots containing data information. The fact check date is based on a combination of self-reported dates provided by fact checkers as part of the ClaimReview markup schema and the date when the ClaimReview-bearing fact check was encountered\footnote{The dates provided in the dataset release are based purely on the self-reported dates from the ClaimReview schema.}. 

\begin{figure}[h]
    \centering
    \includegraphics[width=1\linewidth]{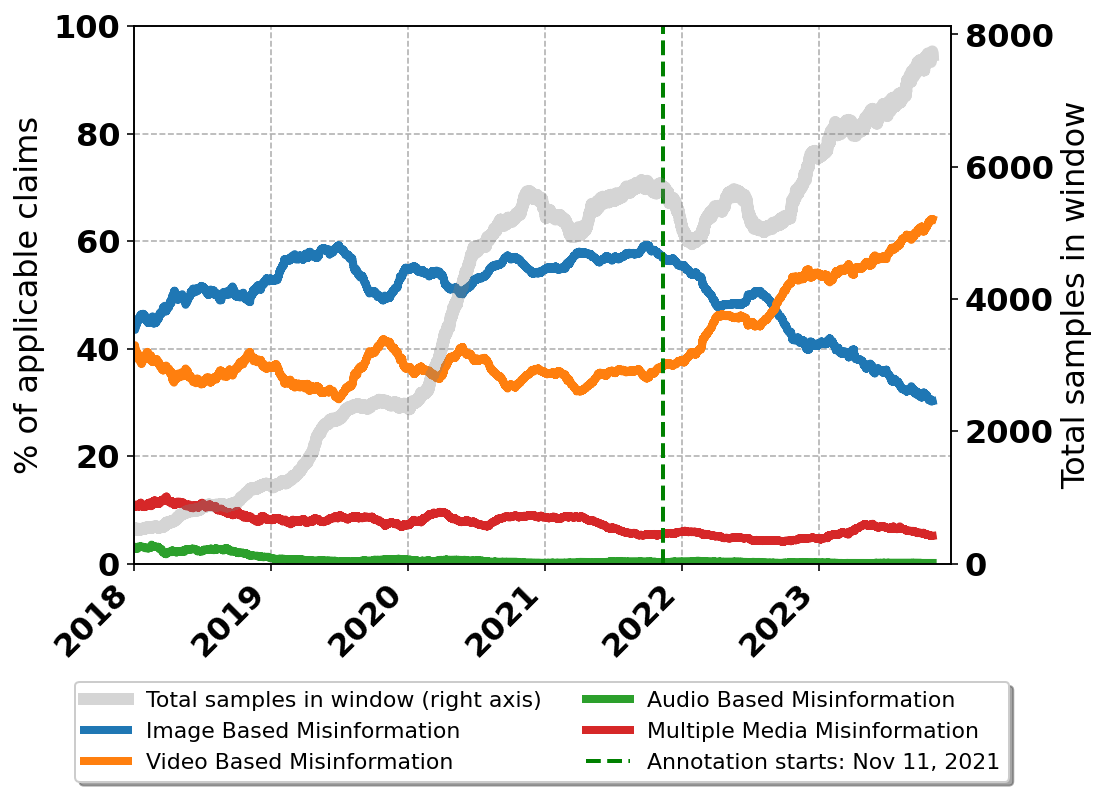}
	\caption{\textbf{Media prevalence by modality in media-based misinformation claims}. Applicable claims are those rated in Stage 1 as relying on media of some form. Note that videos with misinformation-relevant audio but no visual content are considered to be audio-based misinformation. Plot created as in \cref{fig:perc_media_based_claims}.}
\label{fig:misinfo_media_modality}
\end{figure}

The modality of media among misinformation claims is not temporally stable. While previously dominated by images, misinformation claims addressed by fact checks have increasingly involved video on a proportional basis, as seen in \cref{fig:misinfo_media_modality}. This is likely due to the increasing popularity of, and adaptation of fact checkers to, online video as a modality generally. Misinformation claims that rely on multiple media are becoming slightly less common over time, an effect that may also be driven by changes in how online content is encountered and consumed.

\begin{figure}[h]
    \centering
    \includegraphics[width=1\linewidth]{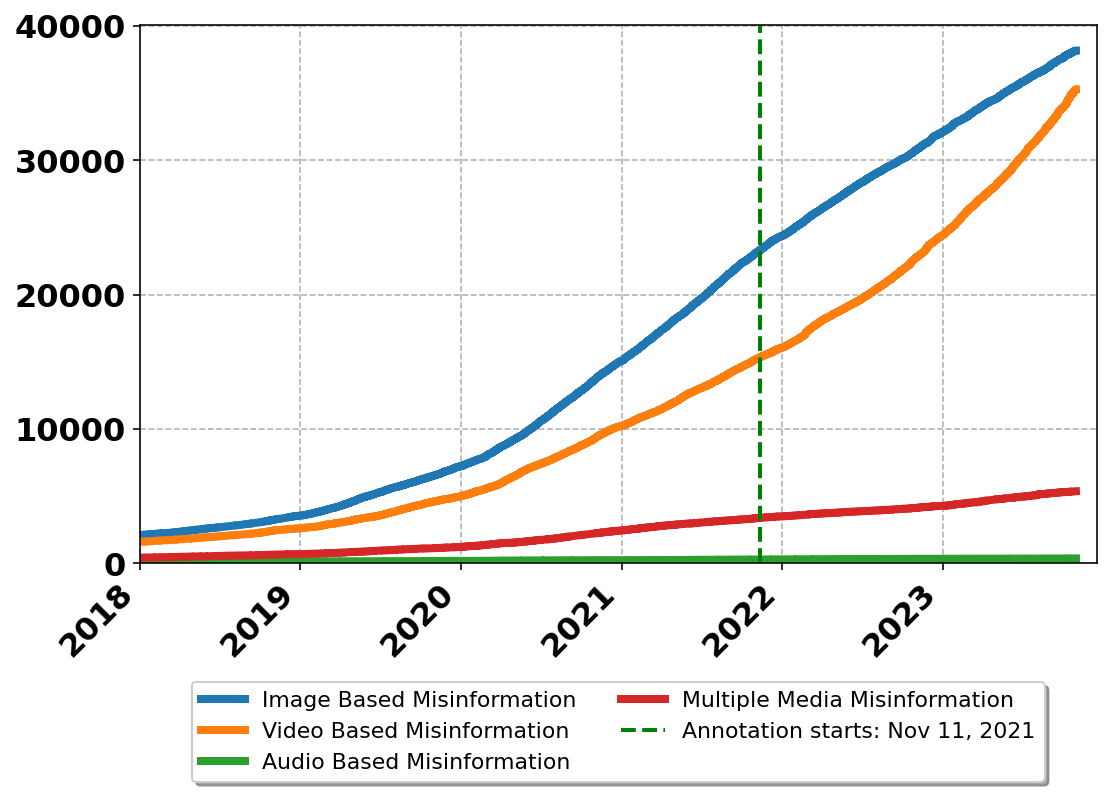}
	\caption{\textbf{Cumulative counts of media-based misinformation claims by media modality}.}
\label{fig:media_type_cumul}
\end{figure}

The secular increase in video-based misinformation is most apparent as a windowed percentage of total media-based misinformation addressed in fact checks over time. Viewed cumulatively in \cref{fig:media_type_cumul} the rise in video-based misinformation does not appear to occur alongside a similarly strong deceleration in other modalities. This suggests increases in fact checker capacity is preferentially allocated to video-based misinformation. 

\begin{figure}[h]
    \centering
    \includegraphics[width=0.8\linewidth]{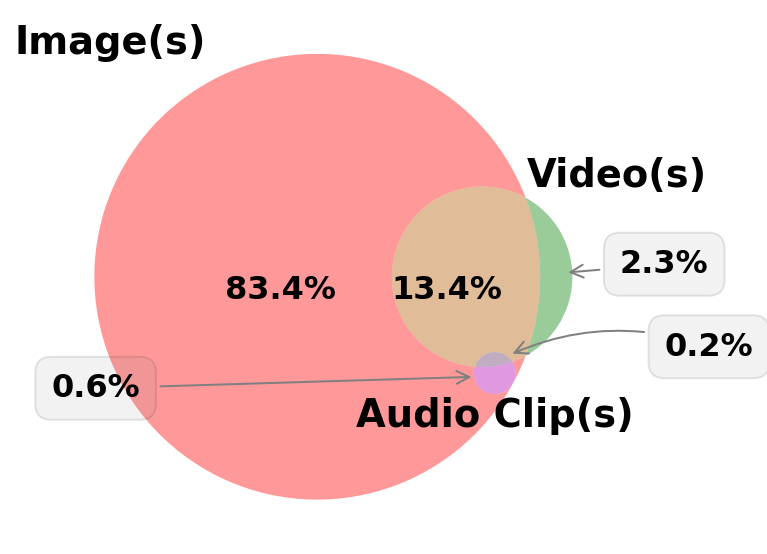}
	\caption{\textbf{Media modality co-occurrence in multiple media misinformation claims}. The co-occurrence of media modalities is visualized as a Venn diagram.}
\label{fig:multi_media_venn}
\end{figure}

The co-occurrence of media modality in multiple-media settings, where misinformation claims depend on more than one media item simultaneously, is visualized in \cref{fig:multi_media_venn}. Images are far more dominant than other modalities in cases where multiple media are involved in a misinformation claim relative to claims that rely on a single media item. In misinformation claims with at least one image present, there are on average 3.0 images (median: 2.0). When videos are present, an average of 2.1 are included (median: 1.0). Video is the next most common modality, and co-occurs with images about 85\% of the time; it's likely that images are used to highlight or point out particular portions of video. Misinformation claims with multiple videos and no images are rare.

\subsection{Image Type Prevalence} \label{sec:image_type_prevalence}

\begin{figure}[h]
    \centering
    \includegraphics[width=1\linewidth]{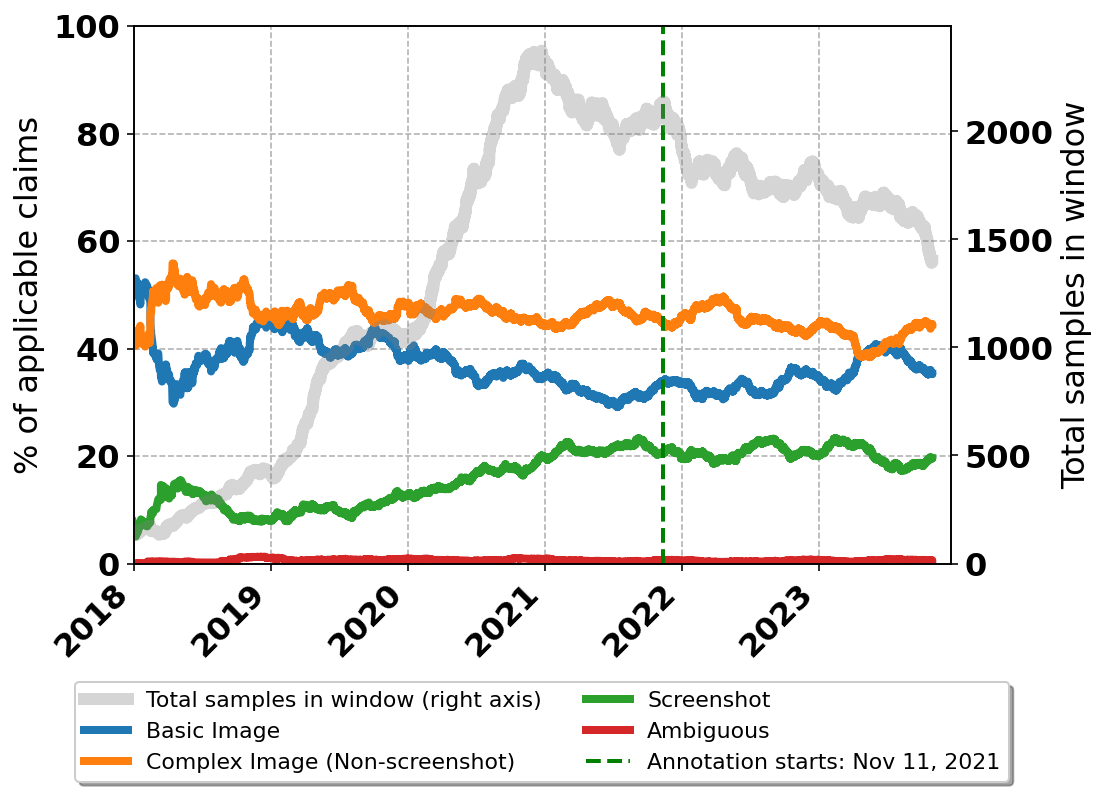}
	\caption{\textbf{Image type in image-based misinformation claims}. Percentages are computed as a proportion of (claim, image) pairs. Plot created as in \cref{fig:perc_media_based_claims}.}
\label{fig:image_type_prevalence}
\end{figure}

Image types (\cref{sec:typology_image_types}) present in image-based misinformation claims are plotted in \cref{fig:image_type_prevalence} and are largely stable. Complex images are consistently more common than basic images. This is unsurprising for several reasons, for example, resharing of content is pervasive online, and during this process images are more likely to transition from being basic to complex via annotation or screenshotting. If driving engagement is a goal of those posting these images, it's also likely that they may take steps to make the images more salient or shareable, which can cause them to become complex images depending on the modifications performed. \cref{fig:image_type_prevalence} breaks screenshots out separately from their parent category, complex images. When taken together, complex images dominate over basic images in proportional terms. A recent bump in simple images is observable in 2023, which is attributable to the rise in the use of generative AI in misinformation claims (see below); such synthesis methods almost always yield images that would be classified as basic.

Note that for \cref{fig:image_type_prevalence} and subsequent plots that represent populations of images, images are deduped using a combination of visual semantic content and text content (when text is present in the image), to prevent over-representation of (image, claim) pairs that are addressed by multiple fact checks. Over-representation is quantified, and regarded as a measure of fact checker interest below (\cref{fig:hype_index} \cref{sec:content_manip}).

\begin{figure}[h]
    \centering
    \includegraphics[width=1\linewidth]{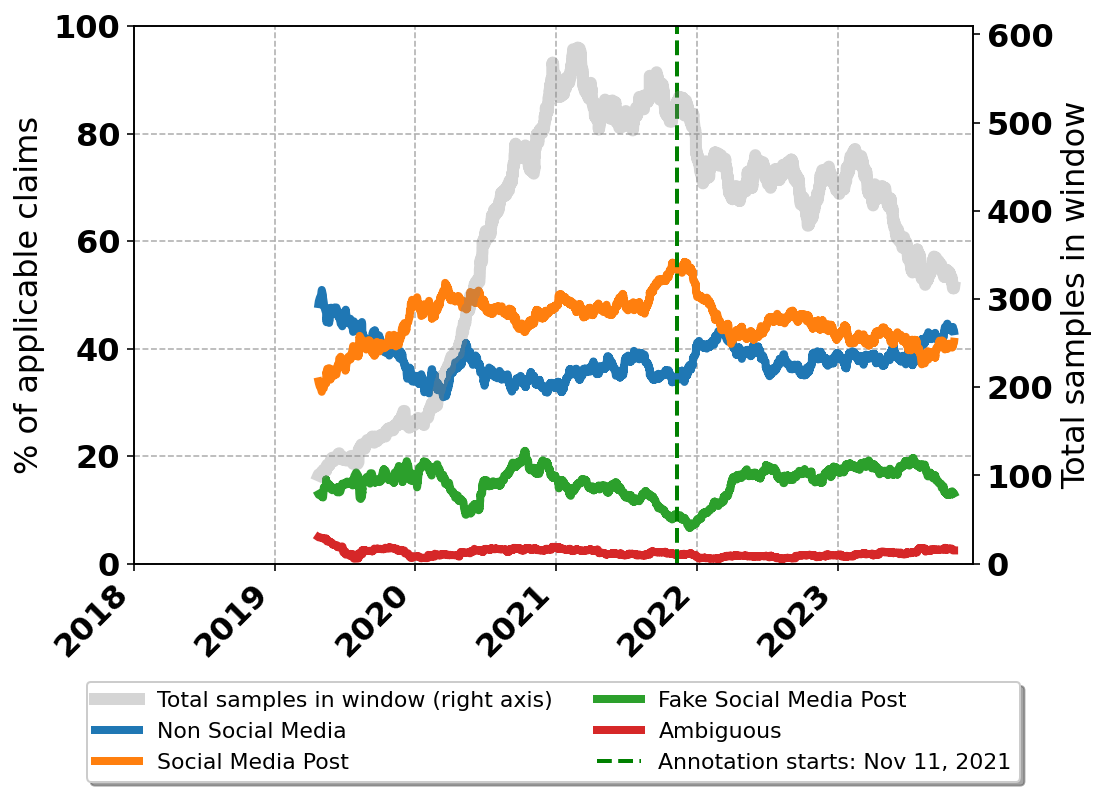}
	\caption{\textbf{Screenshot types in image-based misinformation claims featuring screenshots}. Percentages are computed as a proportion of (claim, image) pairs in misinformation claims featuring screenshots. Plot created as in \cref{fig:perc_media_based_claims}. Because windows containing fewer than 100 examples are not plotted, the left side of the plot is empty.}
\label{fig:screenshot_types}
\end{figure}

Screenshots are further decomposed into three categories (\cref{sec:typology_screenshots}). Their prevalence is plotted in \cref{fig:screenshot_types}. Because of the relatively stringent criteria for inclusion in the screenshot category, estimates of the prevalence of screenshots in misinformation claims are almost certainly an underestimate. We note that screenshots of social media posts are especially common, despite most social media platforms making re-sharing easy. It's not clear what drives this effect, but it may be due to the relative immutability of screenshots: while reposts will reflect edits and policy enforcement made on the original, screenshots will not. Non-social media screenshots are frequently of fake, mocked-up news articles (which may or may not have source identifiers); these are persuasive for the same reasons fake official documents (\cref{sec:fake_official_document}) are: the implication they derive from an authoritative source. In at least some cases, screenshots preserved the pre-correction versions of \emph{real} news articles that contained errors, effectively attaching the authority of a news source to the incorrect claim but omitting any subsequent oversight from that news source.

\subsection{Manipulation Types} \label{sec:manip_types}

Images deployed in misinformation claims are further characterized by \emph{manipulation type}, which indicates the specific steps that were taken to cause the image to present or support misinformation (\cref{sec:typology_manip_types}).

\begin{figure}[h]
    \centering
    \includegraphics[width=1\linewidth]{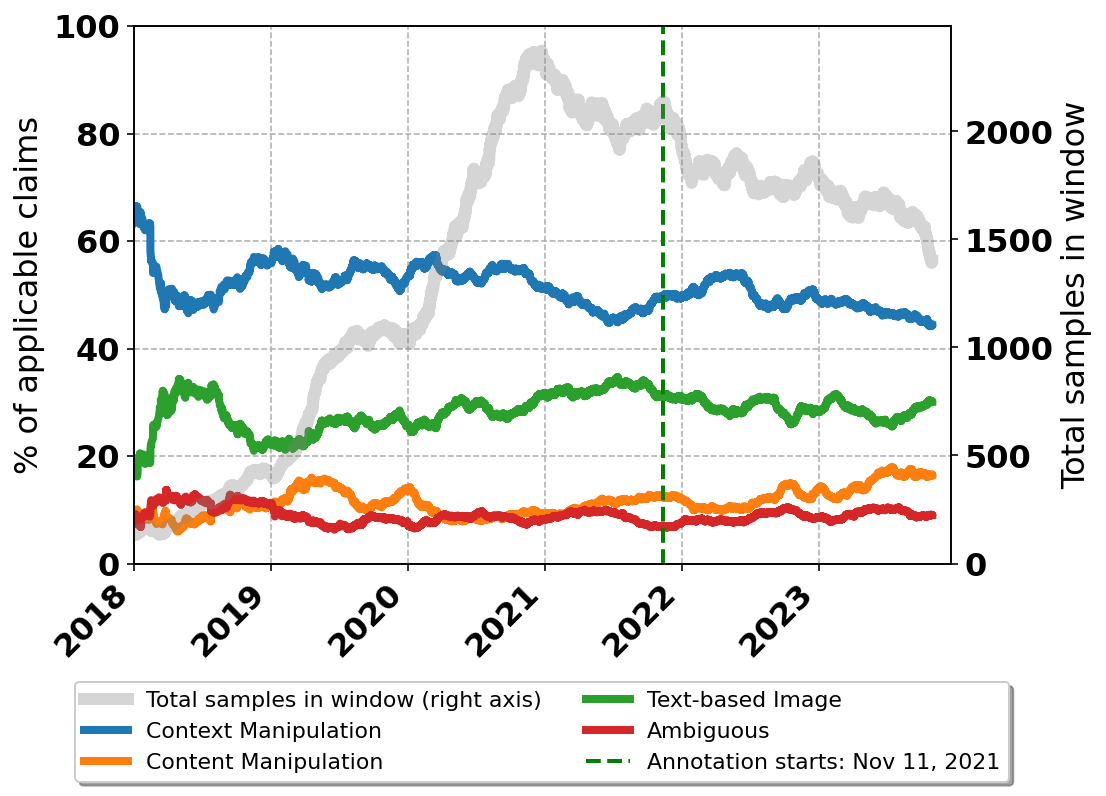}
	\caption{\textbf{Image manipulation types in image-based misinformation}. Percentages are computed as a proportion of (claim, image) pairs where a manipulation has taken place. Plot created as in \cref{fig:perc_media_based_claims}.}
\label{fig:manipulation_types}
\end{figure}

The prevalence of these manipulation categories are plotted in \cref{fig:manipulation_types}. Context manipulations dominate, and have for all points with sufficient data. The respective manipulation prevalence accord with the relative ease of creating the manipulations: content manipulations are the least prevalent and require the use of either image synthesis or image editing techniques; next are text-based images, which nonetheless require some means of rendering text as or over an image, and finally context manipulations, which require no image operations at all, and can be realized entirely by adding a misleading caption. 

Our study provides no measure of the comparative effectiveness or impact of these methods. Despite this, we do not regard the simplicity of producing a particular manipulation as a measure of its ineffective or benign nature. Rather, the consistent popularity of context manipulation suggests their enduring effectiveness as a means of making a compelling false claim.

\subsection{Content Manipulations} \label{sec:content_manip}

\begin{figure}[h]
    \centering
    \includegraphics[width=1\linewidth]{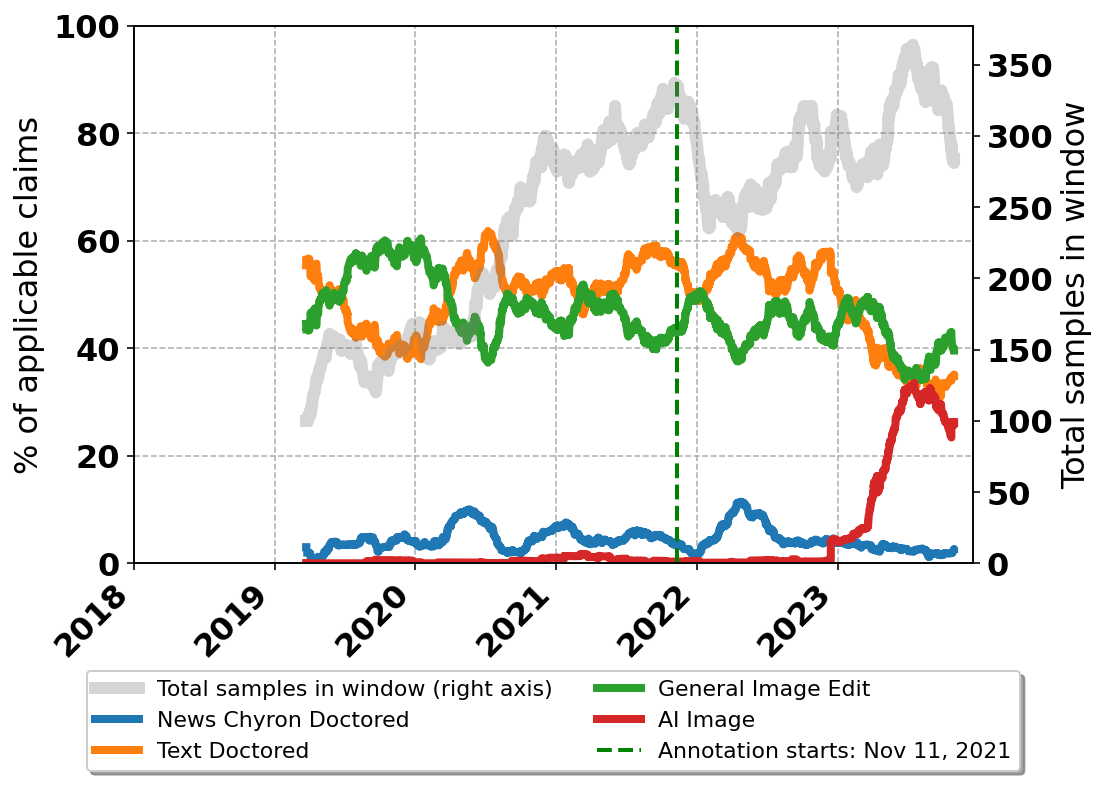}
	\caption{\textbf{Prevalence of content manipulation types as a function of overall content manipulations}. Percentages are computed as a proportion of (claim, image) pairs where a content manipulation has taken place. Plot created as in \cref{fig:perc_media_based_claims}. Because windows containing fewer than 100 examples are not plotted, the left side of the plot is empty.} 
\label{fig:content_manipulation_prev}
\end{figure}

The prevalence of content manipulation sub-types (\cref{sec:typology_content_manip}), as measured against the proportion of content manipulation types overall, is shown in \cref{fig:content_manipulation_prev} and has remained largely stable up until 2023. General (\emph{e.g.} so-called ``photoshops'') and text (applied specifically edits made to text that appears to occur on an object in the scene) manipulations are similarly common and make up a significant majority. Chyron manipulations are a relatively small proportion overall. AI-generated images made up a minute proportion of content manipulations overall until early last year. Starting shortly before 2023, generative AI-generated images begin to rapidly rise as a proportion of overall fact-checked image manipulations; such that the aggregate population of AI-generated images is now far greater than that of manipulated chyrons and nearly as common as text and general content manipulations. Note that the accuracy of the total count of fact-checked AI-content is subject in part to the completeness of their corresponding fact checks (\cref{sec:lims_fut_source_data}): if the fact check does not indicate it's AI, it will not be recorded as such by the rater.

Generative AI models are typically not trained to generate images that convincingly resemble screenshots, memes, or similar image types, and as such generally emit images in the ``basic image'' category (\cref{sec:typology_basic_images}). Accordingly, we see a small increase in the prevalence of basic images overall that correlates well with increasing AI-generated images (\cref{fig:content_manipulation_prev}). Interestingly, the rise of AI images did not produce a bump in the overall proportion of misinformation claims that depend on images (\cref{fig:misinfo_media_modality}) during this period, and image-based misinformation continued to decline on a relative basis as video-based misinformation grew. 

\begin{figure}[h]
    \centering
    \includegraphics[width=1\linewidth]{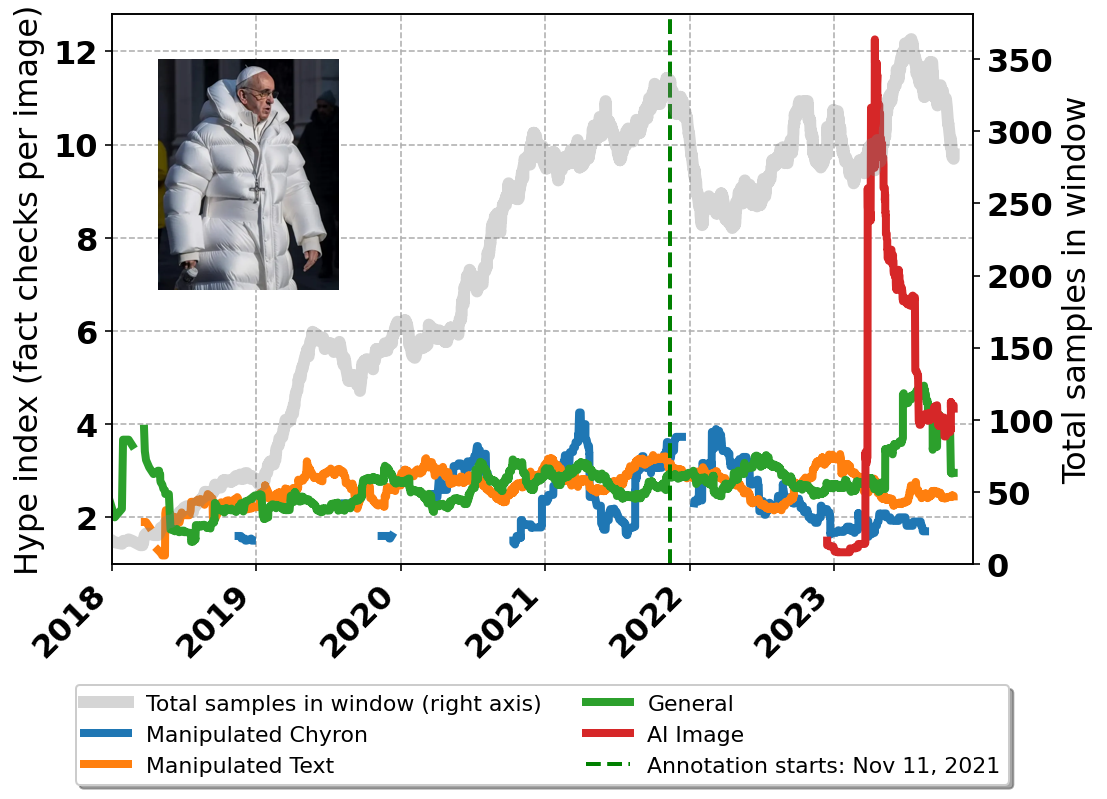}
	\caption{\textbf{The ``Hype Index'' among content manipulations types in image-based misinformation over time}. The ``Hype Index'' is computed as the mean number of fact checks per image over time, separated by content manipulation type. Inset is an AI-generated image of Pope Francis in a white jacket, which went viral in March of 2023 and was subject to numerous independent fact checks as well as being widely featured in the press in general. We interpret this the number of fact checks an image is subject to as an indicator of the fact checker community's view of the importance of fact checking it. Under this interpretation, the massive spike in the index for AI-generated content indicates that fact checkers viewed communicating the nature of these images was of particular public interest (or at least, \emph{of interest} to the public) in early 2023. Plots are otherwise computed as in \cref{fig:perc_media_based_claims}, although the minimum number of datapoints per interval required for inclusion has been changed from 100 to 10 to highlight the sharp spike in fact checks per AI image.} 
\label{fig:hype_index}
\end{figure}

The rise of AI generated images is likely due to a combination of two factors:  growth in the underlying population (as generation tools became more widely known and available in 2023) and a change in attention among fact checkers. The latter can be approximately measured by computing the trend in the number of fact checks addressing a particular image (separated by the content manipulation type), a measure we (somewhat glibly) term the ``Hype Index,'' which is visualized in \cref{fig:hype_index}. 

The Hype Index shows a very large spike in coverage of misinformation claims involving AI-generated images by fact checkers starting shortly after 2023. This coincides well with a spate of viral generative AI images such as the ``Puffy Pope Jacket'' image, inset in \cref{fig:hype_index}. We also note a small (and delayed) bump of interest in general content manipulations and speculate that this may be due to ``sympathetic'' response in interest in such manipulations generally. It may also be due to fact check coverage of generative AI content without the fact check explicitly declaring it as such (more detail in \cref{sec:lims_fut_source_data}), since raters are instructed to label manipulations based exclusively on information present in the fact check. As measured by the index, duplicate fact checker coverage of AI-generated content rapidly decayed and is now at approximately the same level as general content manipulations. This likely represents both the fact checking and online community adjusting in some sense to the increased presence of such images online. The distribution of hype index values for different manipulation types is plotted in \cref{fig:hype_index_violin}.

\subsection{Context Manipulations} \label{sec:context_manip}

Context manipulations, which often do not entail any manipulation to the image pixels themselves, fail to capture public attention to the degree content manipulations---particularly those performed using AI---do. However, our study indicates that they are far more prevalent as a fraction of fact-checked misinformation claims, and likely misinformation overall. Their popularity is probably due to several factors, including the relative ease of creating them and their effectiveness despite their simplicity. 

\begin{figure}[h]
    \centering
    \includegraphics[width=1\linewidth]{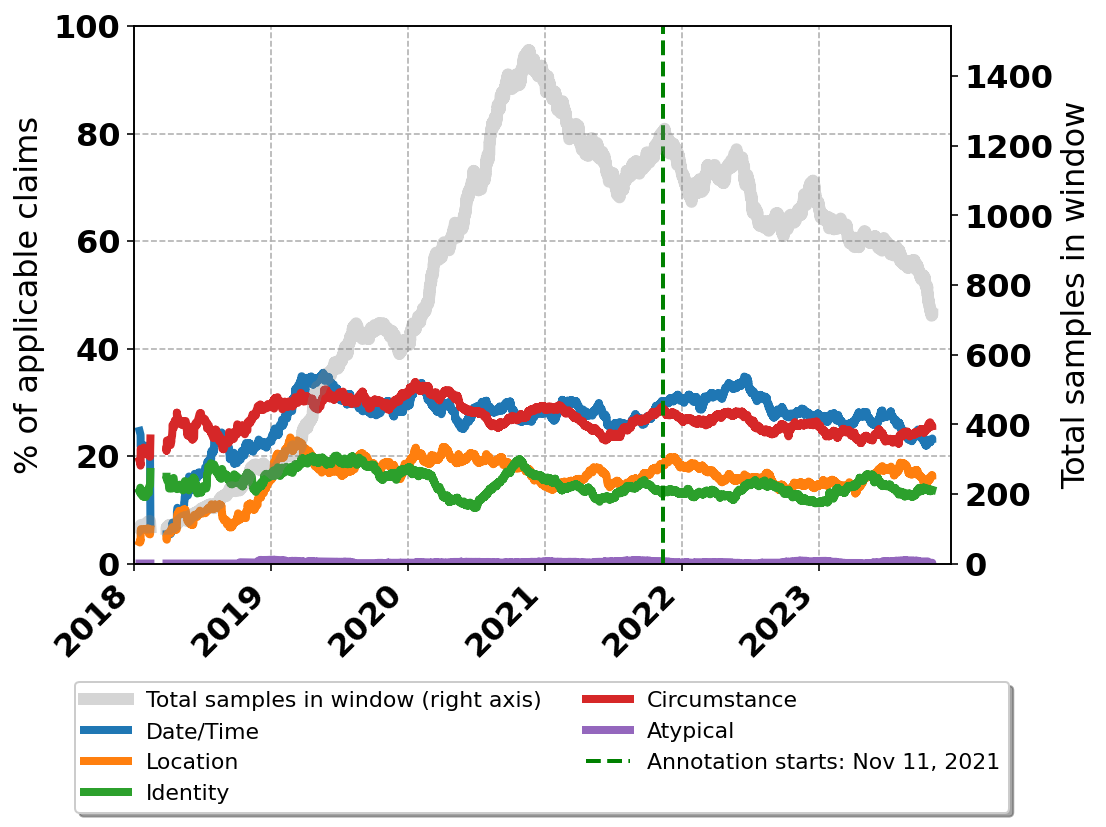}
	\caption{\textbf{Context manipulation types in image-based misinformation with context manipulations}. Percentages are computed as a proportion of (claim, image) pairs where a context manipulation has taken place. Plot created as in \cref{fig:perc_media_based_claims}. Categories are not mutually exclusive. Approximately 40\% of context manipulations do not fall into one of these categories.}
\label{fig:context_manip_types}
\end{figure}

Images that are asserted by the fact check to be subject to context manipulations but which do not fall into the categories defined in \cref{sec:typology_context_manip} are not further characterized, and account for about 40\% of all context manipulations. This reflects relatively stringent criteria for inclusion in a subcategory, and the breadth of what is considered context (more complete categorization of context manipulations is identified as an avenue for further research in \cref{sec:lims_fut_categories}). For example, a context manipulation can falsely state the identity of the \emph{originator} of an image, but not of anything \emph{in} the image, and so it would not fall under the identity manipulation subcategory\footnote{Not capturing these additional subtypes of context manipulation is due mostly to the need to contain the complexity of the overall rater task.}. The prevalence of these categories are shown in \cref{fig:context_manip_types}.

 Unlike content manipulations, there are typically no forensic signals available in the image itself, as it is often un-manipulated. Identifying context manipulations requires identifying inconsistencies among multiple claims pertaining to the image, along with the relative ordering of the claims. Thus, despite the ease with which they can be created, developing mitigations for context manipulation-based misinformation is not straightforward. 

\subsection{Image Text} \label{sec:image_text}

Text is abundant in misinformation-relevant images. The proportion of misinformation-related images bearing text (as detected by optical character recognition) has varied little over time, and hovers at about 80\%. There is considerable, but predictable, variation in the proportion of text-bearing images across categories; text in ``basic image'' (\cref{sec:typology_basic_images}) and AI-generated images (\cref{sec:content_manip}) occurs the most rarely, while text is nearly always present in screenshots (\cref{sec:typology_screenshots}) and ``Text-based'' images (\cref{sec:manip_types}). Further details are available in the appendix (\cref{sec:appx_image_text}). 

\begin{figure}[h]
    \centering
    \includegraphics[width=1\linewidth]{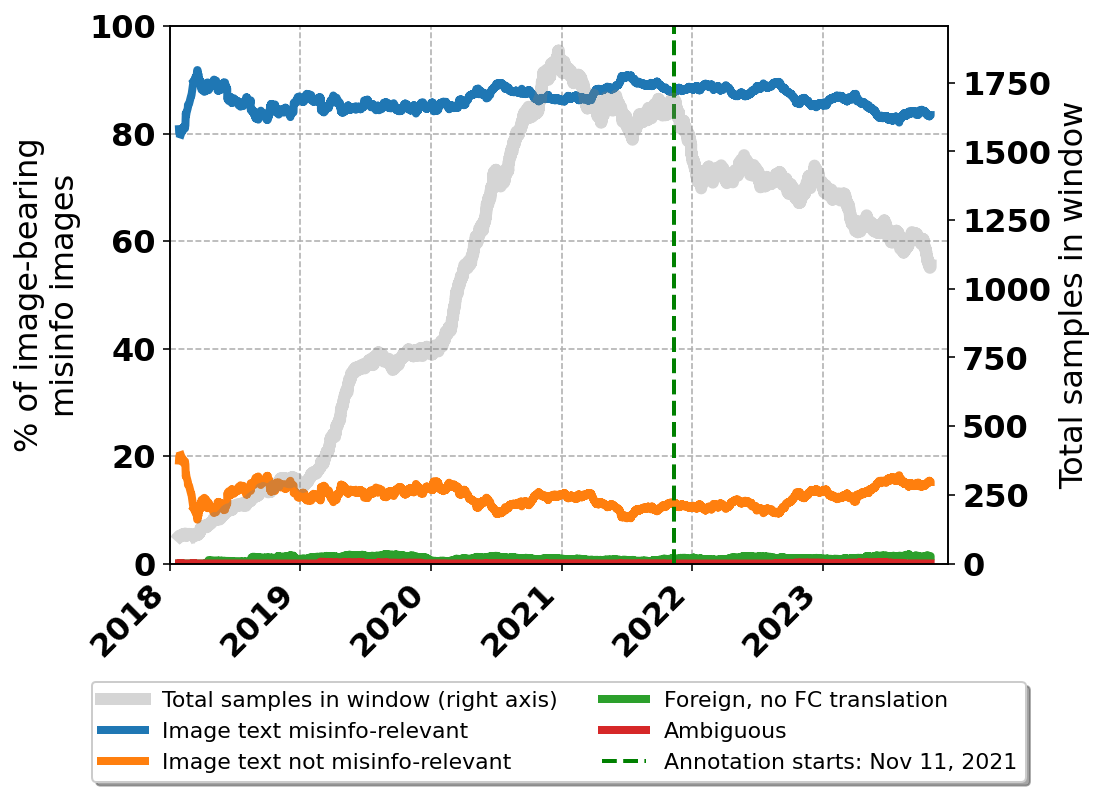}
	\caption{\textbf{Percentage of text-bearing misinformation images with misinformation-relevant text}. Here `FC' is used to abbreviate `fact check.' Plot created as in \cref{fig:perc_media_based_claims}.}
\label{fig:text_relevance}
\end{figure}

Raters noted the presence of text, and did not discriminate between cases where the text occurs on an object in the scene or is digitally overlaid on top of the image. When text was both present and legible, or when a transcript was provided by the fact check (especially in the event the text was in a non-English language), raters were asked to assess whether or not the text was also relevant to the misinformation. Text relevance was determined using criteria similar to those used for the relevance of the image itself: some component of the misinformation claims depends on, or is supported by, the text content in the image. This dependence need not be particularly strong, and the category is used to separate cases where all text that occurs in an image is incidental to the misinformation claim it is associated with. 

Among all annotated misinformation-relevant images bearing text, the proportion where the text is also relevant to the misinformation is plotted in \cref{fig:text_relevance}. As shown in the plot, text---when it occurs---is usually relevant to the misinformation claim. 

\begin{figure}[h]
    \centering
    \includegraphics[width=1\linewidth]{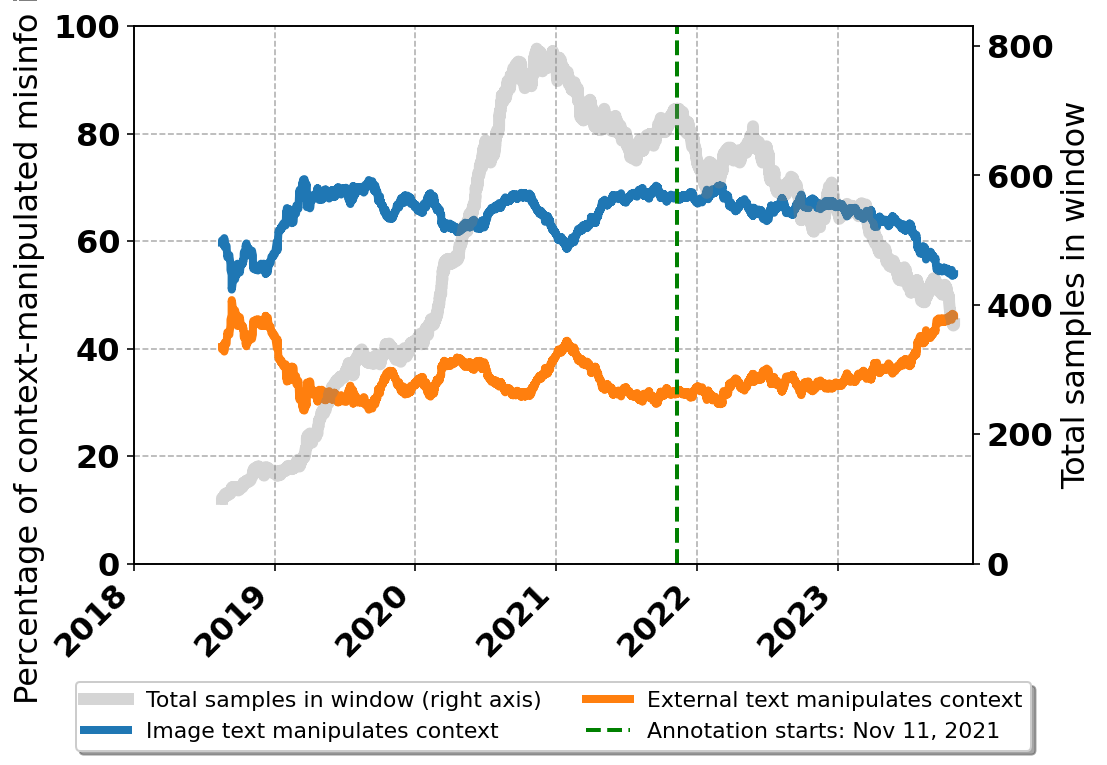}
	\caption{\textbf{Self-contextualizing images}. The prevalence of `self-contextualizing images,' images subject to context manipulations where the text articulating the false context is present in the image itself, among the total population of context manipulations. Plot created as in \cref{fig:perc_media_based_claims}.}
\label{fig:self_contextualizing_images}
\end{figure}

The intersection of image text and context manipulations is of particular interest here. We term cases where the image text provides the contextual claim ``self-contextualizing images'' (perhaps more properly ``self-\emph{mis}contextualizing images''). These are rendered misinformation by the presence of a text-based claim that is both \emph{about} and \emph{within} the image and which, directly or by implication, ascribes it a false context (\cref{sec:typology_self_contextualizing_images}). Their frequency as a proportion of this population is plotted in \cref{fig:self_contextualizing_images}. We were surprised to note that such cases comprise the \emph{majority} of context manipulations. These images are highly shareable on social media platforms, as they don't require that the individual sharing them replicate the false context claim themselves: they're embedded in the image. \Cref{sec:appx_self_contextualizing_image} provides further qualification of the category.

\subsection{Analog Gap} \label{sec:analog_gap}

\begin{figure}[h]
    \centering
    \includegraphics[width=1\linewidth]{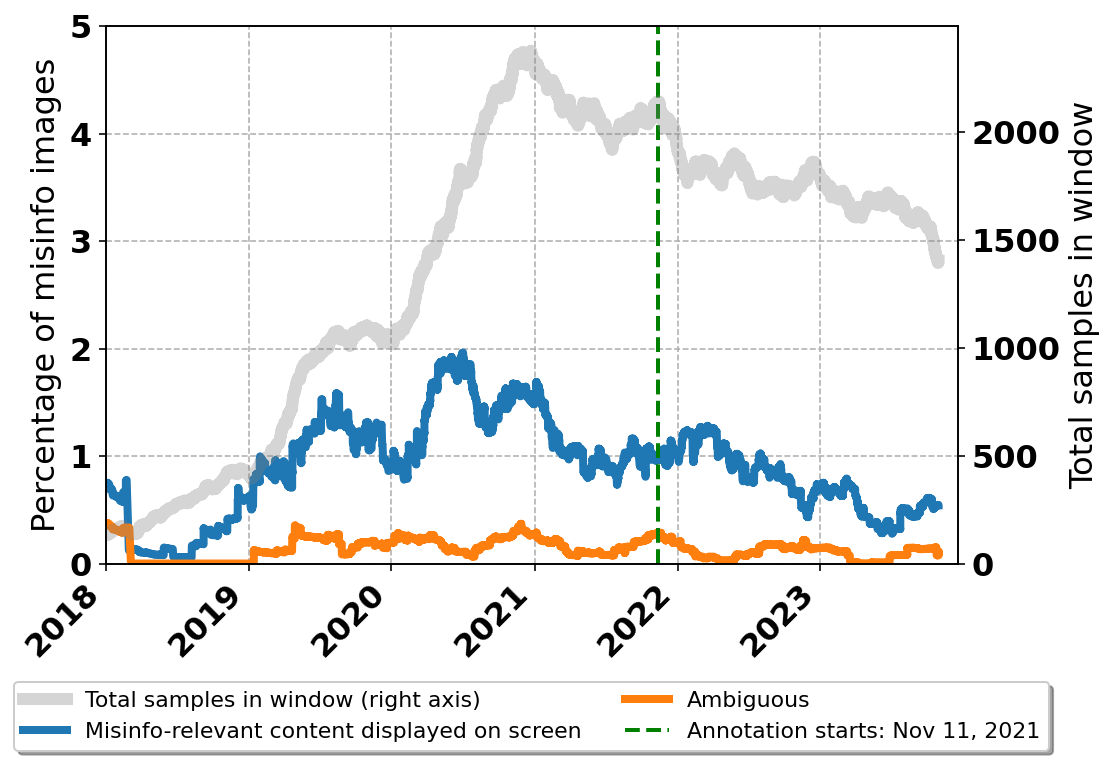}
	\caption{\textbf{Frequency of ``Analog gap'' images}. The proportion of analog gap images, which are characterized by the misinformation-relevant content being displayed on a screen \emph{within} the image itself, is plotted relative to the total population of misinformation-relevant images annotated. Plot created as in \cref{fig:perc_media_based_claims}. Note that the y-axis here is different from other plots: for visibility, the y-axis ranges from 0 to 5\%.}
\label{fig:analog_gap}
\end{figure}

Images exhibiting the so-called ``Analog Gap'' (\cref{sec:typology_analog_gap}) are broken out as a special category. The frequency of such images is plotted in \cref{fig:analog_gap}. While uncommon, these cases do occur. By re-capturing potentially manipulated content, they are especially resistant to forensic analysis, although it's unlikely that this is the intention in most cases. For instance, they may arise simply because a user is unaware of their device's screenshot function, or the display is not directly under their control. Regardless of the intention behind their occurrence, Analog Gap images significantly complicate content analysis. The prevalence of the Analog Gap appears to be undergoing a steady decline in the past several years, which may be due to increased user awareness of screenshot and screen-recording methods, although this is speculation.

\subsection{Fake Official Document}\label{sec:fake_official_document}

\begin{figure}[h]
    \centering
    \includegraphics[width=1\linewidth]{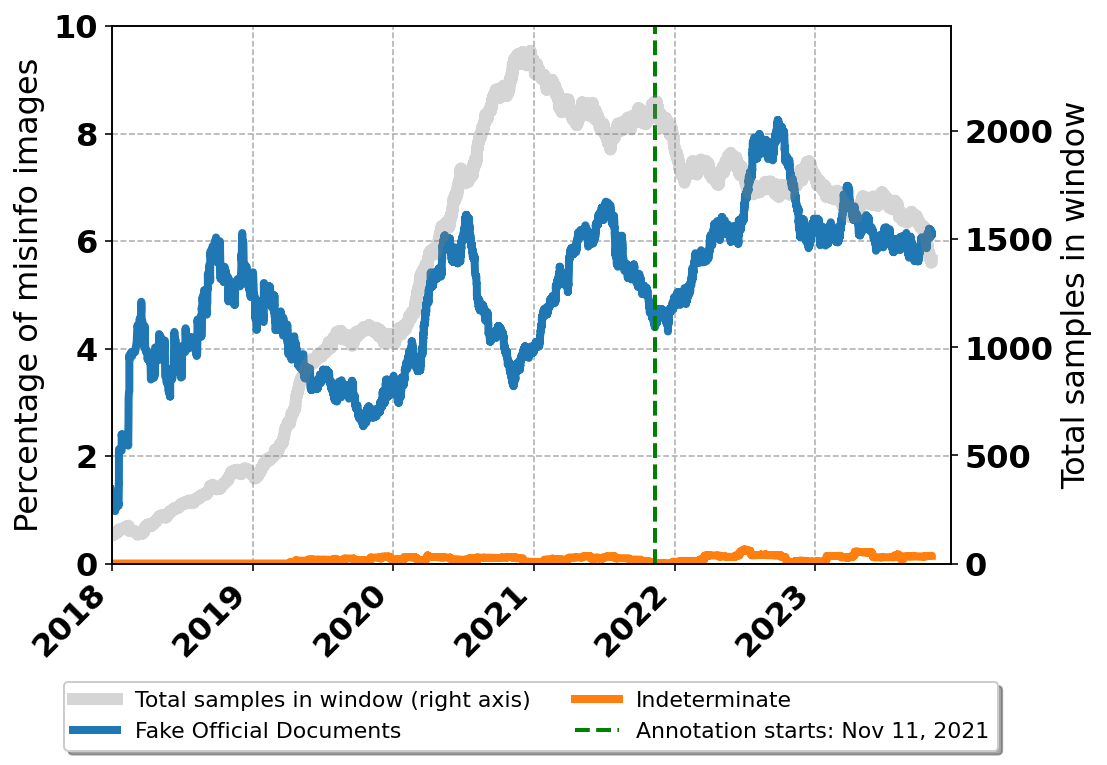}
	\caption{\textbf{Fake Official Documents}. The presence of fake official documents, images appearing to be or contain communications from official organizations, is plotted as a fraction of the total population of misinformation-relevant images annotated. Plot created as in \cref{fig:perc_media_based_claims}. Note that the y-axis here is different from other plots. For visibility, the y-axis ranges from 0 to 10\%.}
\label{fig:false_official_document}
\end{figure}

Fake official documents appear to be or to contain an official communication of some form from a well-known (or made to appear well-known) and reputable organization. The frequency of fake official documents is plotted in \cref{fig:false_official_document}. Such images are of concern as they can disseminate false information during crises, masquerading as an official dispatch containing important information. Particularly salient examples occurred during the COVID-19 pandemic, where misinformation-bearing images would regularly occur appearing to be official information or instruction from a public health body. ``Vaccine exemption cards'' and the like also fall into this category. 

Fake official documents are surprisingly common, and appear to be \emph{increasing} in frequency. They seem to exhibit some cyclicality, although it's not clear if this is being driven by external events or if it is merely noise. This may be due to the wider availability of tools for creating mock documents of a certain type (\emph{e.g.}, fake COVID-19 test results). We explore the possibility that this category depends particularly strongly on the nature of external events in \cref{sec:global_events}.

\subsection{Reverse image search}

\begin{figure}[h]
    \centering
    \includegraphics[width=1\linewidth]{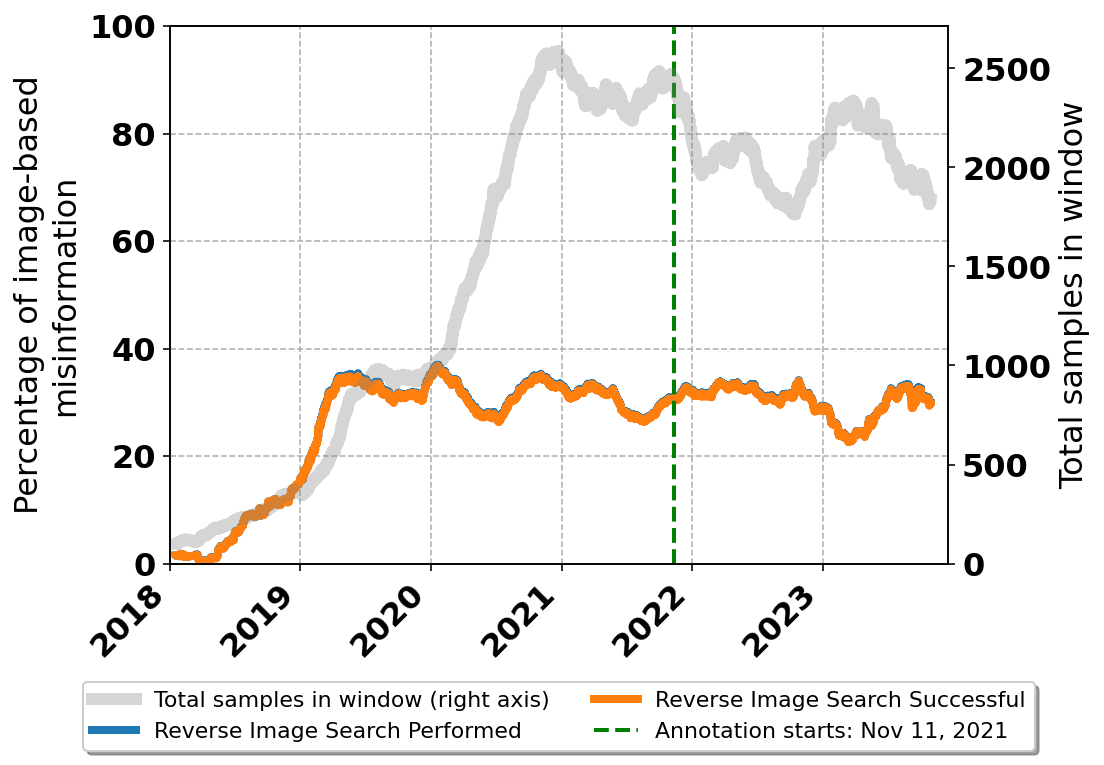}
	\caption{\textbf{Percentage of fact checks reporting reverse-image searches among those that address image-associated misinformation claims}. These proportions are computed with respect to all fact checks that address image-associated misinformation claims. Plot created as in \cref{fig:perc_media_based_claims}. Note that the proportion where a successful search is performed is extremely tightly linked with the proportion where a search is reported at all, making the lines effectively overlaid.}
\label{fig:reverse_image_search}
\end{figure}

Providing an original version of a manipulated image where the manipulation is absent serves as compelling evidence for the manipulation's presence. Such images are provided frequently in fact checks that concern manipulated images. During annotation, raters are asked to capture instances of ``provenance recovery'' by noting cases where the fact check indicates they have conducted a reverse image search, as well as whether or not that search was successful. Our data (visualized in \cref{fig:reverse_image_search}) suggest such reverse image searches are common. Mentions of a successful reverse image search are extremely tightly linked with mentions of conducting a reverse image search at all. This result likely overstates their correlation, though, as it likely represents a so-called ``file-drawer effect'' where unsuccessful searches are omitted from the body of the fact check. 

\begin{figure}[h]
    \centering
    \includegraphics[width=1\linewidth]{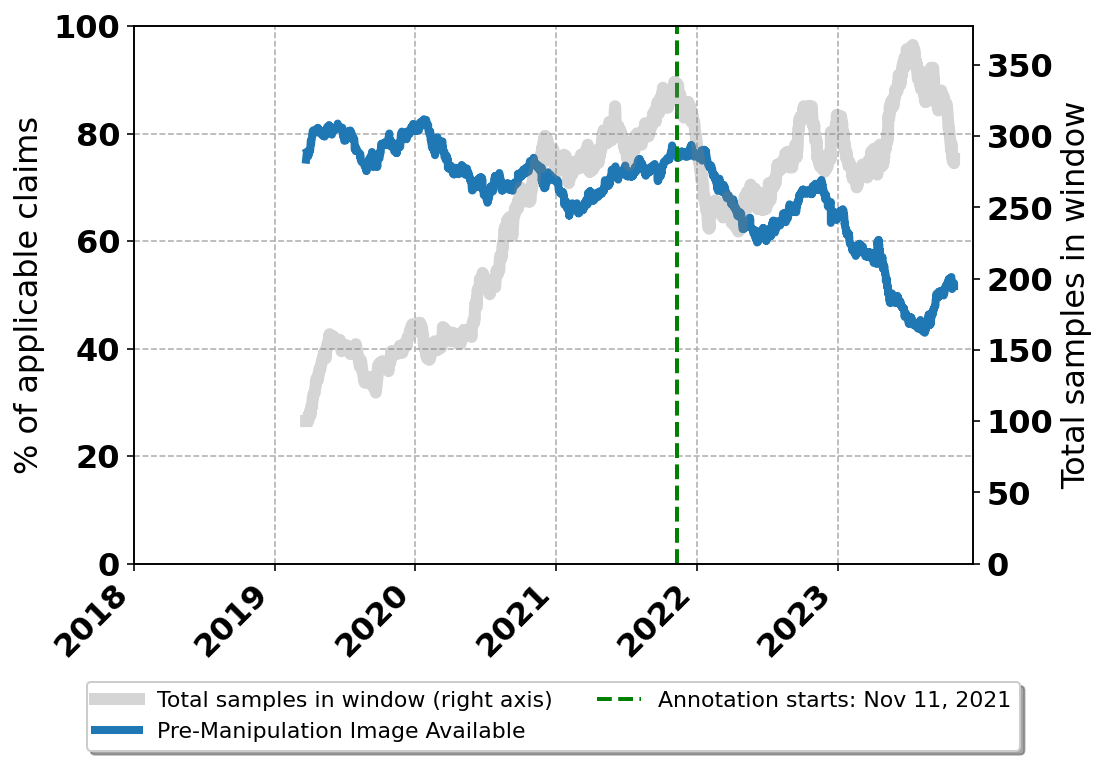}
	\caption{\textbf{Percentage of fact checks presenting a pre-manipulation image when the misinformation claim involves a content-manipulated image}. In the case of content manipulations, fact checks may present an original version of the source image prior to the manipulation taking place as a means of demonstrating its presence. Plot created as in \cref{fig:perc_media_based_claims}.}
\label{fig:pre_manipulation_image}
\end{figure}

In the case of content manipulations, successful provenance recovery is typically accompanied with the presentation of a pre-manipulation version of the image in question. Raters noted the presence of pre-manipulation images in the fact checks, which is plotted in \cref{fig:pre_manipulation_image}. Interestingly, pre-manipulation images appear to show a downward trend, which becomes particularly pronounced in 2023. This coincides with the rise of AI-generated images according to our survey. AI-generated images do not have pre-existing versions, as they are created out of whole cloth, therefore reverse image searches are not useful for finding originals. This may account for part of the decreased frequency in which an pre-manipulation original image is provided by the fact check.

\subsection{Effects of Global Events}\label{sec:global_events}
The volume of fact checks correlates somewhat with global events. The period between the first recorded instance of COVID-19 in the United States on Jan 20, 2020 and the administration of the first of the vaccines on December 14, 2020 corresponds to the most rapid increase in fact checks since the introduction of ClaimReview. 
To measure the effect on the \emph{distribution} of different media types, we checked a battery of notable events against the observed trends. We find little evidence that the events themselves drive shifts in the composition of the misinformation-associated media population. Rather, trends appear to be driven by other phenomena, like the increasing popularity of video platforms or availability of generative-AI methods. 

In some cases, changes are observed consistent with \emph{a priori} expectations. For instance, the ``Fake Official Documents'' category experiences spikes in prevalence following the deployment of COVID-19 tests and vaccinations. However, these observations are by no means conclusive.

Recent global events are often speculated to be the first that is characterized by the dominance of AI-generated misinformation. This has yet to materialize, at least in terms of the presence of AI-generated media. An event associated heavily with AI-generated media may yet occur, but remains in the future.

%% file: sections/7_conclusion.tex
Misinformation is an exceptionally broad phenomenon, unrestricted by modality or topic. It varies by degree, and the boundary is porous and subjective when it abuts categories like satire. Accordingly, researchers developing methods to mitigate its spread and influence may find reckoning with its online distribution, prioritizing research directions, and acquiring in-the-wild samples challenging. In this study, we attempted to survey a restricted (though still expansive in its own right) subset of online misinformation claims with associated media to gain a greater understanding of population-level dynamics, focusing in particular on images that were materially related to misinformation claims. To overcome the challenges of sampling from this domain, we sample misinformation claims and their associated media through the use of publicly-available fact checks with the ClaimReview markup. The work of fact checkers, who operate effectively as ``first responders'' in the fight against misinformation, is invaluable---both in general and for the purposes of grappling with in-the-wild misinformation. 

At first blush, misinformation is often conceptualized in a sense as essentially incorrect statements about the world which may predispose us to think of it as being rendered either verbally or through text. Recently, though, it's become clear that misinformation often relies on the use of media, like images or video, to articulate or support a false claim. Indeed ``often'' is perhaps an understatement, as we show that about 80\% of fact-checked misinformation claims involve media of some kind in a material way. Research efforts must be cognizant of this, and be conducted with this multi-modality in mind. Images and video occur regularly, with video-based misinfo becoming the most common modality recently, possibly due to the rising popularity of video platforms.

Pixel-based forensic methods are a commonly-cited means of identifying image-based manipulations. These methods typically identify content manipulations through irregularities in low-level image properties, like compression and noise artifacts. They are the subject of considerable research, particularly as generative AI becomes an increasing concern. However, we find image manipulations historically tended to be simple. In particular, context manipulations---where images are paired with false claims about what they depict---are extremely common. Context manipulations often use authentic images and involve no deceptive content manipulations at all, making pixel-based forensic methods ineffective in those cases. This points to a need for approaches that can flexibly incorporate contextual information, and compare it to other versions of the media available online. Identifying such manipulations in general is an open research problem, though it's likely some categories, particularly Date/Time and Location, can be automatically inferred given other documents bearing the image.

\begin{figure}
	\centering
    \includegraphics[width=1\linewidth]{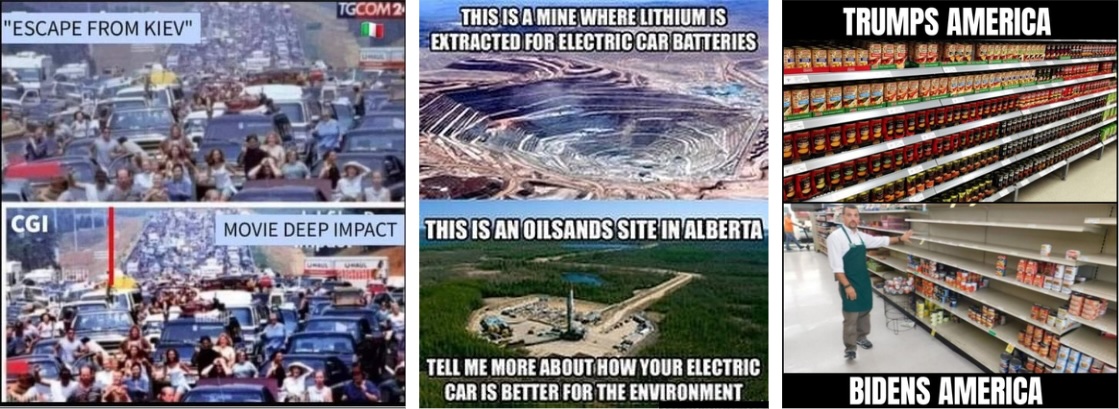}
    \caption{\textbf{Implicit claims in context manipulations}. Images with context manipulations are often not only self-contextualizing, but make the claims implicitly, relying on the viewer making an inference based on the interaction of text and sub-images.} 
    \label{fig:text_image_interactions}
\end{figure}

Context manipulations are not always explicit. Frequently, they're made by implication (or by functioning as ``dog whistles''), not only manipulating the context of the image but also requiring context of their own to apprehend their meaning. These manipulations may also rely on the interaction of multiple sub-images alongside text, including text in the image itself (examples of this are shown in \cref{fig:text_image_interactions}). This complexity underscores a crucial fact: the simplicity of a manipulation does not imply simplicity of detection.

AI-generated content is a prominent topic in discussions of online misinformation. Our data show that, historically, this was not reflected in the online population of misinformation claims, as AI-generated content was extremely rare, at least among claims that are fact-checked. This changed abruptly in 2023, where AI-generated images began to rise rapidly as a fraction of overall rated images, roughly in line with the viral popularity of the ``puffy jacket'' image of the Pope. While AI-generated images did not cause content manipulations to overtake context manipulations, our data collection ended in late 2023 and this may have changed since. Regardless, generative-AI images are now a sizable fraction of all misinformation-associated images.

The sudden prominence of AI-generated content in fact checked misinformation claims suggests a rapidly changing landscape. However, this is occurring alongside already-existing forms of media-based misinformation: the existence of AI-generated content does imply other forms are less effective or less widely used.

%% file: sections/8_limitations_future.tex
Our study was designed with comprehensiveness in mind, with the goal of providing as accurate a representation of the underlying population of media used in-the-wild in misinformation as possible. However, the work is by no means a complete accounting of online media-based misinformation, for reasons beyond the brute facts of rater fallibility or the incomplete availability of information. Below, we outline some of the limitations of this study, and highlight directions for further research.

\subsection{Reliance on Fact Checks}\label{sec:lims_fut_source_data}

Because we rely on fact-checked misinformation claims, we cannot assert that the population level dynamics described in this study exactly correspond to the empirical distribution of misinformation claims overall, only those that are subject to fact checks. However, recognizing the experience and diligence of fact-checking organizations, we regard fact checks as capturing a reasonable sampling of misinformation claims made online and are at least somewhat in proportion to their impact or harm. We therefore presume that the data reflect real trends and phenomena to a meaningful degree.

Raters are instructed to use the fact check as an exclusive source of truth as much as possible for characterizing the image. This approach was adopted to ensure that perspective differences that may be present among the raters are not introduced. This is particularly necessary as misinformation claims are often polarizing. This also means that, in cases where a manipulation has taken place but the fact check does not note its presence\footnote{This was observed in several cases. In particular, some images were noted that bore the hallmarks of AI generation but were not explicitly noted as such.}, the manipulation will not be noted by the rater either. 

\subsection{Language}\label{sec:lims_fut_language}

The fact checks we sampled to measure misinformation claims were English language only. This was motivated by the availability of raters, the need to be able to perform training and the feasibility of quality control. Misinformation is highly topical and depends on cultural context. Understandably, fact checks do not always provide a completely explicit accounting of the claim, and assume some level of fluency on the part of the reader. Therefore, familiarity with the current trends in the population targeted by the misinformation is often necessary to understand the nature of the misinformation claim. 

Nonetheless, the restriction to English language fact checks is a substantial limitation of our work; not only would the topic of misinformation claims be significantly different in other languages, the trends in modality and manipulations would likely differ as well. Because of this, claims about the generality of the trends observed here are not possible; further study is necessary. Until then, the results here will carry a large English-centric caveat.

\subsection{Data Attrition}\label{sec:lims_fut_attrition}

Internet platforms often act to remove or restrict low-quality content, particularly misinformation. Because of this, fact checks experience pronounced ``link rot,'' even if the underlying misinformation claim persists elsewhere.

Because fact checks were assessed in multiple stages in our study at different points in time, we could estimate the rate at which it occurs by finding cases where the link identified by the rater in one stage fails to work in subsequent stages. We find attrition occurs most rapidly in the time immediately after the fact check is published, followed by a slow decay thereafter, which gradually accelerates over time.   

Attrition is a limitation of this study, as it's unlikely it occurs in a random manner; we expect more virulent or harmful misinformation claims are more likely to undergo attrition than relatively benign examples. This phenomenon is not only a problem for this study, but also in general, as it diminishes the utility of the fact check if it prevents those accessing it from knowing what claims it actually addresses. Many fact checkers use archival sites to overcome this problem, but their use is not universal and their reliability in general is not known.

\subsection{Modality}\label{sec:lims_fut_modality}

Annotation focused heavily on the images (rather than video or audio) that participate in misinformation claims. While the original intent was to apply similarly granular characterization to video and audio content, the complexity of the task and the within-modality diversity of the media required the scope be narrowed.

The use of different manipulation methods almost certainly depends heavily on the modality; with their ease of creation and the ease with which they can be dismissed by viewers figuring heavily into their selection. Further, the typology developed here was shaped in part by the empirical prevalence of those categories, making their applicability outside of the image modality limited, at least until it can be determined that they generalize.

We hope that future research in other modalities can use this work as a starting point.

\subsection{Finer Categorization}\label{sec:lims_fut_categories}

As mentioned above, about 40\% of context manipulations do not fit into the pre-defined subcategories (\cref{sec:context_manip}). The variety of other context manipulation types were apparent during data collection (for example, media creator misattribution). Given the number of other categories we hoped to capture, keeping the task from becoming unreasonably complex was necessary to ensure that raters could maintain accuracy and efficiency. 

To properly study the breadth of context manipulations, subsequent stages would be necessary where raters could apply finer-grained annotations. Further deep dives, particularly in categories that exhibit large within-category variation or with large numbers of examples that defy subcategorization would afford greater understanding.

%% file: sections/appendix.tex
\section{Inclusion Criteria} \label{sec:appx_inclusion_criteria}

Fact checks were eligible for at least some annotation if they had associated ClaimReview markup. During stage one, fact checks that: \begin{itemize}
    \item Addressed more than one claim (``anthology'' fact checks).
    \item Failed to load correctly or were otherwise not inaccessible via a web browser.\footnote{Fact checks were enqueued for annotation in Stage 1 on two occasions, separated in time, to avoid transient loading errors.}
    \item Were not in English.
\end{itemize} were noted as such and not further annotated. 

Lastly, fact checks that did not regard the claim as false, misleading or misinformation were also not further annotated. Fact checking organizations use a variety of heterogeneous labeling systems to indicate veracity. Raters were instructed to err on the side of excluding misinformation claims that were assessed as ``possibly true'' or ``unclear,'' etc.

\section{Media Typology Details} \label{sec:appx_media_typology}

\subsection{Image Types} \label{sec:appx_typology_image_types}

The image type categories were selected to reflect properties of the images that are relevant to an investigator, like a fact checker, engaged in determining its authenticity, as the means used for investigating an image present in a misinformation claim differ based on the nature of the image and the nature of the suspected manipulation. For example, while reverse image search methods are reliable in the case of natural images and are widely deployed in fact checking \cite{10017287}, they may be less useful for images with large amounts of diagrammatic content and text \cite{jones2022abstract}, which contain less information than natural images and bear elements that are frequently and identically repeated across images. Additionally, content may be captured by taking a screenshot and re-shared, sometimes several times. Each of these can introduce graphical elements into the new image, and reduce the fidelity of the underlying original. 

We distinguish between \textbf{basic images}, which depict a unified scene as would be captured by a physical camera, and \textbf{complex images} which have additional elements that are clearly added after-the-fact. \textbf{Screenshots} are a subtype of complex images where GUI elements are unambiguously visible and indicate that the image results from a screenshot.

Images with substantial ``natural'' visual content (as opposed to textual, schematic, symbolic, etc), in particular basic images, can be readily debunked by finding original instances that provide either the correct context (in the case of context manipulations) or a pre-manipulation version (in the case of content manipulations). This is frequently possible as the photograph-like quality of such images makes them well-suited for retrieval by reverse-image search tools. Computational forensic methods generally require images will be of this type, allowing authentic images to be distinguished from manipulated images by identifying irregularities in camera and noise properties. An understanding of their proportion relative to the overall population is therefore relevant to those developing forensic methods. 

Complex images frequently require additional steps during analysis, such as isolating and extracting natural-image sub-components present or conducting a verbatim search for any overlaid text present in the image. Complex images of all kinds may consist of multiple basic images in a mosaic or collage, or contain significant regions without naturalistic visual content, complicating provenance-based investigations.

Screenshots introduce additional considerations. The process of capturing a screenshot may entail the recompressing or resizing the underlying content, which is known to inhibit the efficacy of some forensic methods \cite{Verdoliva_2020}, particularly those that rely on compression artifacts introduced during the process of image manipulation. Finally, GUI elements that can be attributed to specific platforms or applications can narrow the search space for finding originals, as the possible sources are limited relative to the open web. When the screenshot depicts a social media post by a well-known personality, it may be possible to verify their authenticity by simply checking to see if a corresponding post exists on the authentic social media feed.

\subsection{Analog Gap Images} \label{sec:appx_typology_analog_gap}
The ``Analog Gap'' (or ``Analog Hole'', ``Rebroadcast Attack'' \cite{fan2018rebroadcast}) was originally identified in the context of copy protection systems, where it was considered an unavoidable vulnerability. The gap occurs when media displayed for human consumption is recaptured. Since any method used to prevent copying and transmission of the media must be disabled prior to the audience perceiving it, there is always an opportunity for the media to be re-recorded.

A similar vulnerability exists in media forensics \cite{fan2018rebroadcast}. Forensic analysis methods often exploit subtle statistical properties present in manipulated media that are left behind by the manipulation or synthesis tools used. These signals can be destroyed when the media is recaptured during playback or display, as (1) all the camera properties of the image will appear (and are) original and consistent and (2) the recapturing represents what is effectively an unpredictable and continuous composition of numerous natural and difficult-to-model transformations of the media. Additionally, it makes recovering the provenance of media difficult for similar reasons. In the case of images, this can occur when a photograph is taken of a screen displaying the image. 

\subsection{Manipulation Types} \label{sec:appx_typology_manip_types}

In this study, ``manipulation'' is not used to refer to arbitrary changes made to the image, but have a more specific definition with two important distinctions:

\begin{enumerate}[label=(\roman*)]
    \item Manipulations are labeled if the manipulation causes the image to create or support the fact checked misinformation claim. An image may be used in a misinformation claim and contain photo-retouching for aesthetic purposes; this only qualifies it as having a content manipulation (\emph{see} \ref{sec:content_manip}) if the content changes are relevant to the misinformation claim. Similarly, if an image is created using image editing software, then posted as satire, and subsequently re-posted without the satirical context, the primary manipulation type is \emph{context manipulation} (as important context has been omitted thereby allowing it to spread misinformation, \emph{see} \ref{sec:typology_context_manip}).
    \item Manipulations need not apply to the image's content or (or file metadata). For example, an image may be associated with a caption that contains a false claim about what the image shows; this would be considered a context manipulation.
\end{enumerate}

The manipulation types are not intended to be exhaustive or complete. Rather, they were developed to capture the type of evidence that could be presented by a hypothetical fact checker to demonstrate the presence of misinformation dispositively; a complex forensic analysis of an image to show that a content manipulation has taken place may not be necessary if an earlier version of an image can be identified that contains, for example, a clear satirical context. Their granularity is also a function of the rater's ability to discriminate between them and the availability of information in typical fact checks.

\subsubsection{Fake Official Documents} \label{sec:appx_typology_false_official_documents}

Fake official documents are created to give the false impression of an official communication from an organization or authority. This category was not present in our original formulation of the typologies, but was added once their prevalence was recognized. These types of images are particularly relevant during public safety events, where communication from officials can be crucial. 

These images are intentionally constructed to appear ``official'' in some sense, and may be created out of whole cloth, or may be modified versions of authentic communications. Features added to an image to make it appear official range from a single logo to highly elaborate, intricately adorned replicas. The originating organization may also be invented, but made to appear (or sound) similar to well-known or authentic groups.

Social media posts falsely associated with accounts representing official organizations or figures are labeled differently and as screenshots. \emph{See} \cref{sec:typology_screenshots}. Were they not fabricated, it would be possible to obtain an equivalent version from the official source. Screenshots or other images of manipulated news broadcasts are not members of this category.

\section{Data Collection Rater Task} \label{sec:appx_dataset_task}

The complexity of the initial task introduced considerable cognitive load on raters, and went through a number of iterations and redesigns before settling on the final version. Data collected before the task was finalized were invalidated \footnote{\emph{All} responses are made available in the dataset release. Those not used in this paper's analysis, either because they were collected before a methodological change was made or were considered otherwise unusable for a variety of reasons, are marked as invalid.}. 

A ``staged'' approach was adopted, where misinformation claims present in fact checks passed from upstream stages to downstream stages according to upstream annotations. In each stage, raters were provided with the fact check along with any relevant information from upstream annotations, and asked to respond to questions further characterizing those misinformation claims and their associated media. The bulk of the questions would only be made visible when previous responses made them relevant and applicable, otherwise they were hidden from view. Questions were typically forced choice, although raters generally had the ability to indicate when they were not confident in their responses. A small number of questions were free text input. The four stages completed are visualized in \cref{fig:task_stages_schematic}.

In all cases, raters were instructed to treat the fact check as a source of truth for annotations, and to refrain from making independent judgments. Exceptions to this are cases where the judgement could be made objectively or concerned categories developed specifically for this study (principally, the distinction between basic and complex images, \emph{see} \ref{sec:typology_image_types}) and never when related to the veracity of the misinformation claim being examined or the evidence establishing this veracity. In cases where misinformation claims are addressed by multiple fact checks presenting conflicting assessments of claims, or raters provided conflicting responses, those conflicts will be reflected in the dataset. 

\subsection{Stage 1: Rough Misinformation Characterization}\label{sec:appx_stage_1}

All misinformation claims received preliminary annotations in Stage 1, in which raters determined whether or not the misinformation claim relied on the presence of media and the modality of those media. Raters also sought to identify a URL (present in the fact check) containing the misinformation claim as it existed originally online, either its source in cases where it was still available online, or as part of an archived webpage created by the fact checker. When media was involved in the misinformation claim and materially relevant to it (because it supports, provides evidence for, or directly contains the misinformation claim), the URL provided must contain that media. In addition to characterizing the modality of the media media used, raters also indicated whether or not multiple media were used in the claim (i.e., several images, a video alongside stills, etc., \emph{see} \ref{sec:typology_media_based}). 

A total of 337,723 annotations were collected in Stage 1, of which 88,430 were discarded for various reasons (significant changes to task guidance, raters were found to be too inaccurate, etc), resulting in a total of 249,293 annotations. As measured by fact check URL, claims had an average annotation replication of 1.8. Replications were not evenly distributed, as claims that were randomly selected to undergo quality control to evaluate rater accuracy had higher replication; 46.3\% of claims had more than 1 Stage 1 replication. A total of 135,838 claims were annotated in Stage 1, although occasionally claims could not be re-annotated after data were discarded due to the fact check no longer being available; 24 claims were dropped in this manner.

Stage 1 annotation began on November 11, 2021, although the first response not discarded was collected on November 23, 2021. The last stage 1 annotation was performed on October 31, 2023.

\subsection{Stage 1M: Multiple-media Misinformation Claims}\label{sec:appx_stage_1m}

Misinformation claims rated as depending on multiple media in Stage 1 (\emph{see} \cref{sec:appx_stage_1}) were annotated in Stage 1M, which further refined the characterization of the nature of the media used and the number of distinct media items. Raters were presented with a link to the fact check and the original source (as identified in Stage 1) and any images associated with the misinformation claim as identified in previous iterations of Stage 1M annotation of a given claim.

A total of 46,752 annotations were collected in Stage 1M. 9,213 were discarded, resulting in a final total of 37,539 annotations. As measured by fact check URL, claims had a mean replication of 7.3. High replication was present in Stage 1M because multiple media misinformation claims typically involve at least one image. In such cases, where images were among the multiple media involved, raters were asked to identify them one by one, until raters indicated that all images had been identified. 98.1\% of all misinformation claims participating in Stage 1M had more than one replication. 5,133 misinformation claims were identified as involving multiple media and passed to Stage 1M, although 21 were lose due to annotations being discarded and the corresponding fact checks being unavailable 

Stage 1M annotation began on June 14, 2023 and ended on November 17, 2023. 

\subsection{Stage 2: Fine-Grained Annotation of Single-Image Misinformation Claims}\label{sec:appx_stage_2}

In cases where misinformation claims depended on a single image (according to Stage 1), fine-grained annotation was performed in Stage 2. Raters were presented with the URL to the fact check, the URL to the original source (as identified in Stage 1), and the misinformation-relevant image identified in Stage 1. Stage 2 is applied independently for each (claim, image) pair identified in Stage 1. 

Stage 2 contains dozens of categories, most centering on the manipulation present in, or applied to, the image to support or make the misinformation claim (does the image contain AI-based manipulation, did the fact check conduct a reverse-image search, etc). Unlike Stages 1 and 1M, very few annotations were discarded (a total of 69). Each (claim, image) pair has an average of 2.6 Stage 2 annotations, and 84.9\% of claims have more than 1 replication.

Stage 2 was collected between Feb 23, 2023 and Nov 17, 2023. 

\subsection{Stage 2M: Fine-Grained Annotation of Images from Multiple-Media Misinformation Claims}\label{sec:appx_stage_2m}

Stage 2M was the final stage conducted, and is applied to misinformation claims bearing multiple media including at least one image. For every image associated with such a claim, a Stage 2M annotation is applied. Raters are presented with a fact check URL and an original source as identified in Stage 1. Images associated with the claim, identified in Stage 1M (\emph{see} \cref{sec:appx_stage_1m}), are also shown; one image is the "primary" image, with the remainder secondary. The primary image is the main subject of annotation.

All categories present in Stage 2 are also present in Stage 2M. Additionally, several categories extended to the other images presented, with the forced choice having roughly the form:
\begin{itemize}
  \item The primary image shown is of Type \emph{X}.
  \item The primary image is not of Type \emph{X}, but another image shown is of Type \emph{X}.
  \item No image presented is of Type \emph{X}.
\end{itemize}
where raters were instructed to select the first applicable response.

Because the raters are instructed to use the fact check as a source of truth (instead of making their own judgements), this is intended to capture cases where the fact check indicates a manipulation of some type is present but does not specify the images to which it is applied.

A total of 21 annotations were discarded in Stage 2M. Each (claim, image) pair had an average of 1.9 annotations, and 68.2\% of (claim, image) pairs had more than one replication.

\section{Hype Index} \label{sec:appx_hype_index}

\begin{figure}[h]
    \centering
    \includegraphics[width=1\linewidth]{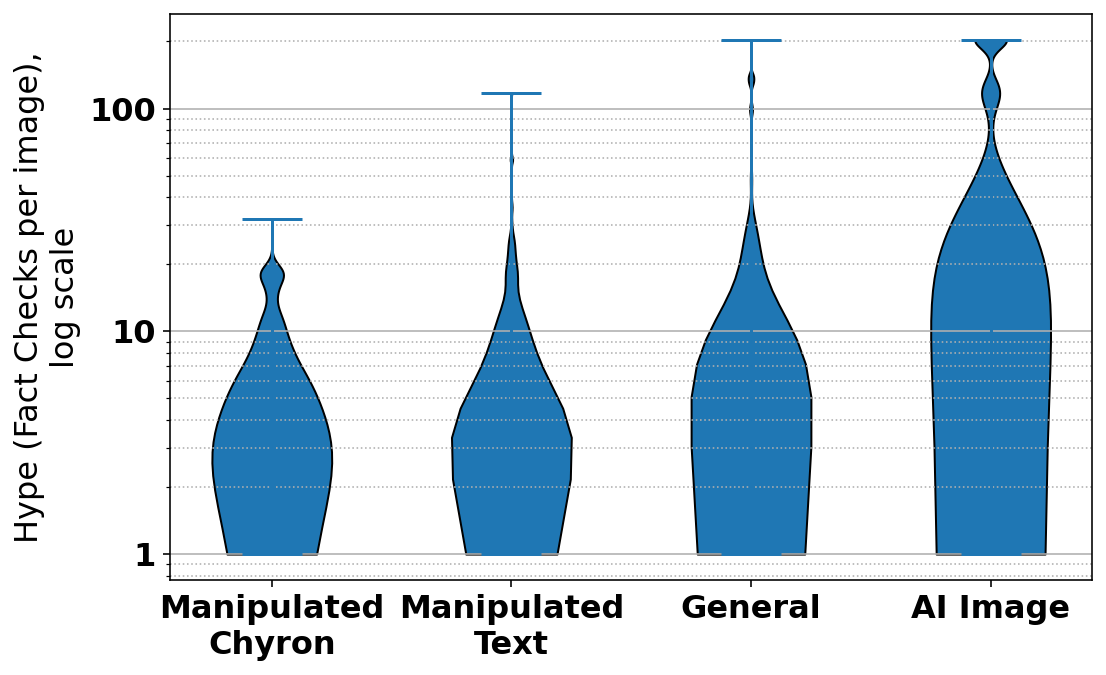}
	\caption{\textbf{Distribution of hype by content manipulation type}. Here, the value of the hype index is plotted on the y axes, and the number of instances of a given type with that hype index value is represented by the width of the plot at that point.}
\label{fig:hype_index_violin}
\end{figure}

We visualize "hype" for a particular image-based misinformation category by counting the number of independent fact checks that address that image, under the presumption that this is a loose measure of interest among fact checkers, which in turn we presume is proportional to their assessment of the fact check's utility to the public. This is visualized over time in \cref{fig:hype_index}. The distribution of hype index values over all instances, grouped by type, is shown in \cref{fig:hype_index_violin}. AI-image hype is robustly greater than other categories across multiple instances. 

\section{Image Text} \label{sec:appx_image_text}

\begin{figure}[h]
    \centering
    \includegraphics[width=1\linewidth]{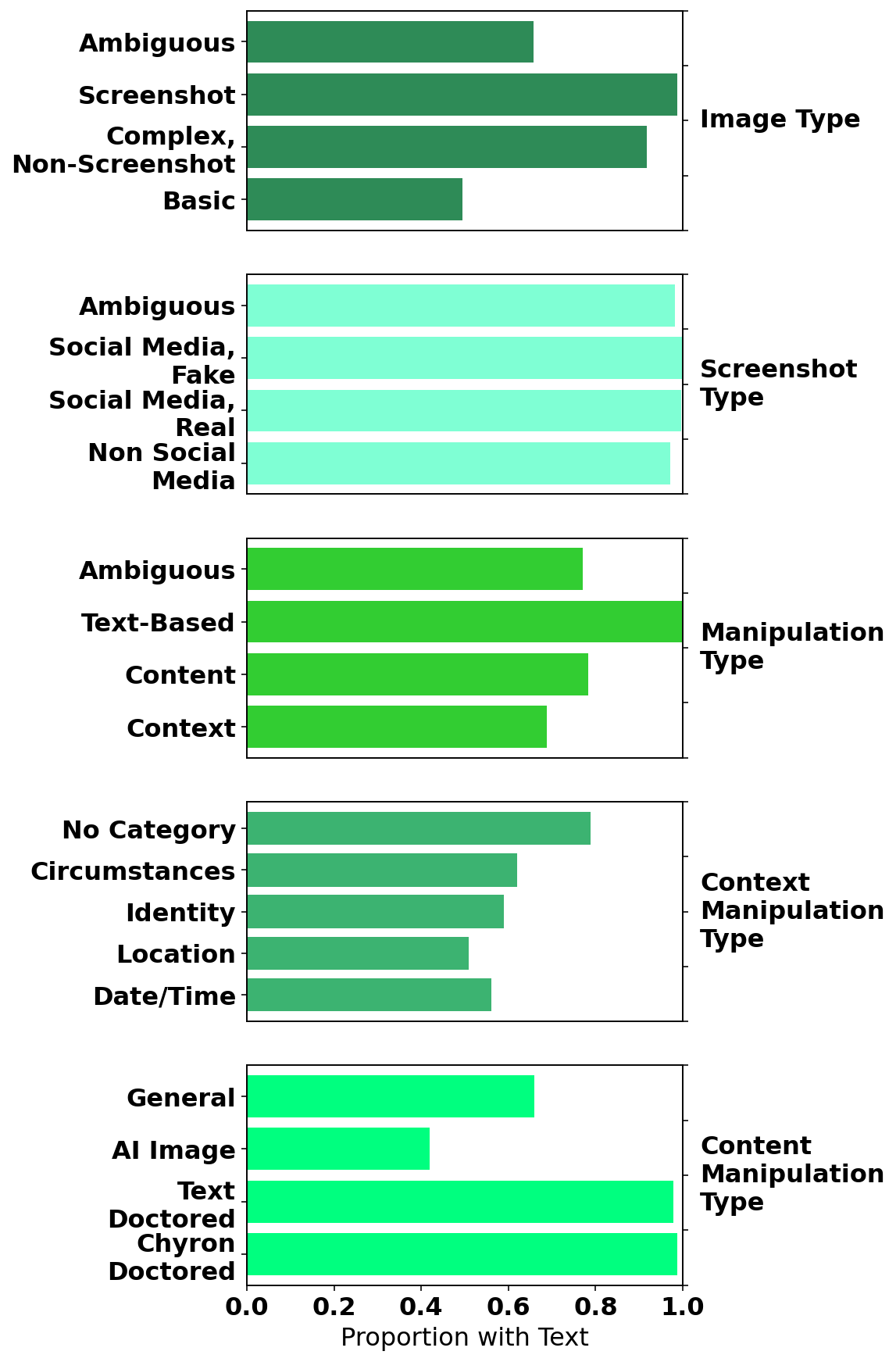}
	\caption{\textbf{Text presence by image category}. The proportion of images containing text of any kind within each category, as recognized by optical character recognition, is plotted on the x-axis.}
\label{fig:text_presence}
\end{figure}

\begin{figure}[h]
    \centering
    \includegraphics[width=1\linewidth]{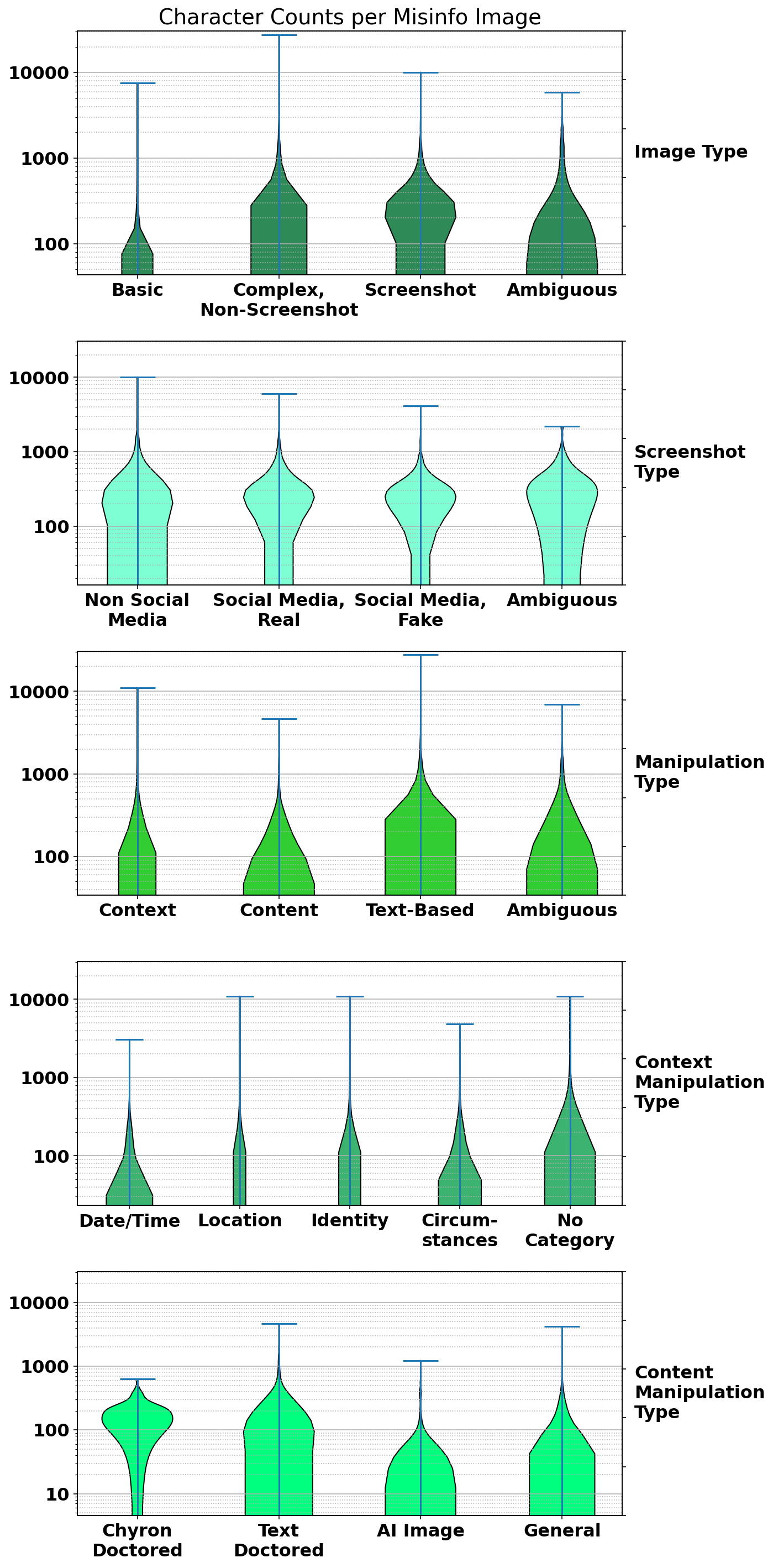}
	\caption{\textbf{Distribution of the number of text characters by image category}. The number of characters per image was computed using optical character recognition. The distribution of counts across instances within a particular category is realized as a violin plot.}
\label{fig:character_counts_violin}
\end{figure}

A majority of images annotated in this study contained text of some kind, either present in the scene on objects or overlaid digitally on the image itself. Using OCR, the proportion of images bearing text of some form, by image category, is displayed in \cref{fig:text_presence}. The distribution over the counts of text characters within images on a per-class basis is shown in \cref{fig:character_counts_violin}. 

\section{Self-contextualizing image} \label{sec:appx_self_contextualizing_image}

\begin{figure}[h]
    \centering
    \includegraphics[width=1\linewidth]{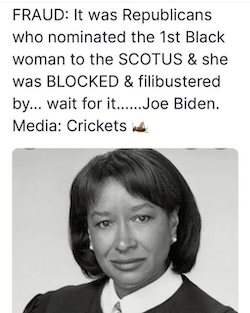}
	\caption{\textbf{Example of weakly self-contextualizing image}. Self-contextualizing images (\emph{see} \cref{sec:image_text}) are context-manipulated images where the text articulating the false context is present in the image itself. This is a broad category; the above image is included as an example of images with this annotation.}
\label{fig:self_contextualizing_image_example}
\end{figure}

The population of self-contextualizing images depicted in \cref{fig:self_contextualizing_images} may be inflated relative to the strict sense of what constitutes a context manipulation by the relative broadness of the category's criteria. \cref{fig:self_contextualizing_image_example} is an example of this broadness; the image itself does not directly support the image being made. However, the contextual implication made by the text present is that this is a portrait of a Supreme Court Justice nominee whose nomination was blocked. In reality, it is circuit judge Janice Rogers Brown, who was considered but not nominated for the Supreme Court. While this form of weak interaction between text and image content is considered a valid context manipulation, the claim is understood in all cases to not directly relate to what is depicted in the image.

%% file: main.bbl
\begin{thebibliography}{10}

\bibitem{midjourney_kaggle}
Midjourney user prompts \& generated images (250k).
\newblock
  \url{https://www.kaggle.com/datasets/succinctlyai/midjourney-texttoimage}.
\newblock Accessed: 2024-03-20.

\bibitem{StateoftheFactCheckersReport2023}
State of the fact-checkers report 2023.
\newblock Technical report, International Fact-Checking Network, 2023.

\bibitem{allcott_et_al_2019}
Hunt Allcott, Matthew Gentzkow, and Chuan Yu.
\newblock Trends in the diffusion of misinformation on social media.
\newblock {\em Research \& Politics}, 6(2):2053168019848554, 2019.

\bibitem{cosmos}
Shivangi Aneja, Christoph Bregler, and Matthias Nie{\ss}ner.
\newblock Catching out-of-context misinformation with self-supervised learning.
\newblock {\em CoRR}, abs/2101.06278, 2021.

\bibitem{Birdsell1996}
David~S. Birdsell and Leo Groarke.
\newblock Toward a theory of visual argument.
\newblock {\em Argumentation and Advocacy}, 33(1):1--10, 1996.

\bibitem{brennen2020types}
J~Scott Brennen, Felix~M Simon, Philip~N Howard, and Rasmus~Kleis Nielsen.
\newblock {\em Types, sources, and claims of COVID-19 misinformation}.
\newblock PhD thesis, University of Oxford, 2020.

\bibitem{chen2008determining}
Mo~Chen, Jessica Fridrich, Miroslav Goljan, and Jan Luk{\'a}s.
\newblock Determining image origin and integrity using sensor noise.
\newblock {\em IEEE Transactions on information forensics and security},
  3(1):74--90, 2008.

\bibitem{chen2023twigma}
Yiqun Chen and James Zou.
\newblock Twigma: A dataset of ai-generated images with metadata from twitter,
  2023.

\bibitem{chesney2019deep}
Bobby Chesney and Danielle Citron.
\newblock Deep fakes: A looming challenge for privacy, democracy, and national
  security.
\newblock {\em Calif. L. Rev.}, 107:1753, 2019.

\bibitem{diresta2024spammers}
Renee DiResta and Josh~A Goldstein.
\newblock How spammers and scammers leverage ai-generated images on facebook
  for audience growth.
\newblock {\em arXiv preprint arXiv:2403.12838}, 2024.

\bibitem{fan2018rebroadcast}
Wei Fan, Shruti Agarwal, and Hany Farid.
\newblock Rebroadcast attacks: Defenses, reattacks, and redefenses.
\newblock In {\em 2018 26th European Signal Processing Conference (EUSIPCO)},
  pages 942--946. IEEE, 2018.

\bibitem{grinberg2019fake}
Nir Grinberg, Kenneth Joseph, Lisa Friedland, Briony Swire-Thompson, and David
  Lazer.
\newblock Fake news on twitter during the 2016 us presidential election.
\newblock {\em Science}, 363(6425):374--378, 2019.

\bibitem{hameleers2020picture}
Michael Hameleers, Thomas~E Powell, Toni~GLA Van Der~Meer, and Lieke Bos.
\newblock A picture paints a thousand lies? the effects and mechanisms of
  multimodal disinformation and rebuttals disseminated via social media.
\newblock {\em Political communication}, 37(2):281--301, 2020.

\bibitem{ps_battles}
Silvan Heller, Luca Rossetto, and Heiko Schuldt.
\newblock The ps-battles dataset - an image collection for image manipulation
  detection.
\newblock {\em CoRR}, abs/1804.04866, 2018.

\bibitem{unesco_ipsos_survey_2023}
{Ipsos/UNESCO}.
\newblock Survey on the impact of online disinformation and hate speech, 2023.

\bibitem{jones2022abstract}
Shawn~M. Jones and Diane Oyen.
\newblock Abstract images have different levels of retrievability per reverse
  image search engine, 2022.

\bibitem{kff_survey_2023}
{KFF}.
\newblock Kff misinformation poll snapshot: Public views misinformation as a
  major problem, feels uncertain about accuracy of information on current
  events, 2023.

\bibitem{10017287}
Sohail~Ahmed Khan, Ghazaal Sheikhi, Andreas~L. Opdahl, Fazle Rabbi, Sergej
  Stoppel, Christoph Trattner, and Duc-Tien Dang-Nguyen.
\newblock Visual user-generated content verification in journalism: An
  overview.
\newblock {\em IEEE Access}, 11:6748--6769, 2023.

\bibitem{li2020picture}
Yiyi Li and Ying Xie.
\newblock Is a picture worth a thousand words? an empirical study of image
  content and social media engagement.
\newblock {\em Journal of Marketing Research}, 57(1):1--19, 2020.

\bibitem{mccrae2022multi}
Scott McCrae, Kehan Wang, and Avideh Zakhor.
\newblock Multi-modal semantic inconsistency detection in social media news
  posts.
\newblock In {\em International Conference on Multimedia Modeling}, pages
  331--343. Springer, 2022.

\bibitem{meeker2016internet}
Mary Meeker.
\newblock Internet trends 2016, 2016.

\bibitem{facebook_widely_viewed_content_2023_Q3}
Meta.
\newblock Facebook widely viewed content report: Q3 2023.
\newblock Technical report, November 2023.
\newblock Accessed: April 1, 2024. Downloaded from archive:
  https://transparency.fb.com/data/widely-viewed-content-report?gk_enable=stc_nov_2023\#prior-reports.

\bibitem{nakamura2020rfakeddit}
Kai Nakamura, Sharon Levy, and William~Yang Wang.
\newblock r/fakeddit: A new multimodal benchmark dataset for fine-grained fake
  news detection, 2020.

\bibitem{newman2023misinformed}
Eryn~J Newman and Norbert Schwarz.
\newblock Misinformed by images: How images influence perceptions of truth and
  what can be done about it.
\newblock {\em Current Opinion in Psychology}, page 101778, 2023.

\bibitem{nielsen2022mumin}
Dan~Saattrup Nielsen and Ryan McConville.
\newblock Mumin: A large-scale multilingual multimodal fact-checked
  misinformation social network dataset, 2022.

\bibitem{paris2019deepfakes}
Britt Paris and Joan Donovan.
\newblock Deepfakes and cheap fakes.
\newblock 2019.

\bibitem{ap_norc_survey_2023}
{Pearson Institute/AP-NORC}.
\newblock The american public views the spread of misinformation as a major
  problem, 2021.

\bibitem{posetti2018short}
Julie Posetti and Alice Matthews.
\newblock A short guide to the history of ‘fake news’ and disinformation.
\newblock {\em International Center for Journalists}, 7(2018):2018--07, 2018.

\bibitem{reis2020dataset}
Julio~CS Reis, Philipe Melo, Kiran Garimella, Jussara~M Almeida, Dean Eckles,
  and Fabr{\'\i}cio Benevenuto.
\newblock A dataset of fact-checked images shared on whatsapp during the
  brazilian and indian elections.
\newblock In {\em Proceedings of the international AAAI conference on web and
  social media}, volume~14, pages 903--908, 2020.

\bibitem{Rossler_2019_ICCV}
Andreas Rossler, Davide Cozzolino, Luisa Verdoliva, Christian Riess, Justus
  Thies, and Matthias Niessner.
\newblock Faceforensics++: Learning to detect manipulated facial images.
\newblock In {\em Proceedings of the IEEE/CVF International Conference on
  Computer Vision (ICCV)}, October 2019.

\bibitem{thorne}
James Thorne, Andreas Vlachos, Christos Christodoulopoulos, and Arpit Mittal.
\newblock {FEVER:} a large-scale dataset for fact extraction and verification.
\newblock {\em CoRR}, abs/1803.05355, 2018.

\bibitem{Verdoliva_2020}
Luisa Verdoliva.
\newblock Media forensics and deepfakes: An overview.
\newblock {\em IEEE Journal of Selected Topics in Signal Processing},
  14(5):910–932, August 2020.

\bibitem{Wang17j}
William~Yang Wang.
\newblock "liar, liar pants on fire": {A} new benchmark dataset for fake news
  detection.
\newblock {\em CoRR}, abs/1705.00648, 2017.

\bibitem{wang2021understanding}
Yuping Wang, Fatemeh Tahmasbi, Jeremy Blackburn, Barry Bradlyn, Emiliano
  De~Cristofaro, David Magerman, Savvas Zannettou, and Gianluca Stringhini.
\newblock Understanding the use of fauxtography on social media.
\newblock In {\em Proceedings of the International AAAI Conference on Web and
  Social Media}, volume~15, pages 776--786, 2021.

\bibitem{wang2023diffusiondb}
Zijie~J. Wang, Evan Montoya, David Munechika, Haoyang Yang, Benjamin Hoover,
  and Duen~Horng Chau.
\newblock Diffusiondb: A large-scale prompt gallery dataset for text-to-image
  generative models, 2023.

\bibitem{wardle2017information}
Claire Wardle and Hossein Derakhshan.
\newblock {\em Information disorder: Toward an interdisciplinary framework for
  research and policymaking}, volume~27.
\newblock Council of Europe Strasbourg, 2017.

\bibitem{weikmann2023visual}
Teresa Weikmann and Sophie Lecheler.
\newblock Visual disinformation in a digital age: A literature synthesis and
  research agenda.
\newblock {\em new media \& society}, 25(12):3696--3713, 2023.

\bibitem{yang2023visual}
Yunkang Yang, Trevor Davis, and Matthew Hindman.
\newblock Visual misinformation on facebook.
\newblock {\em Journal of Communication}, 73(4):316--328, 2023.

\bibitem{zhu2023genimage}
Mingjian Zhu, Hanting Chen, Qiangyu Yan, Xudong Huang, Guanyu Lin, Wei Li,
  Zhijun Tu, Hailin Hu, Jie Hu, and Yunhe Wang.
\newblock Genimage: A million-scale benchmark for detecting ai-generated image,
  2023.

\end{thebibliography}
